\documentclass[manuscript]{acmart}

\AtBeginDocument{%
  \providecommand\BibTeX{{%
    \normalfont B\kfern-0.5em{\scshape i\kern-0.25em b}\kern-0.8em\TeX}}}
    
\setcopyright{acmcopyright}
\copyrightyear{2021}
\acmYear{2021}
\acmDOI{}

\acmJournal{TOSEM}
\acmMonth{10}

\usepackage{graphicx}
\usepackage{xurl}
\usepackage{booktabs}
\usepackage{multibib}
\usepackage{natbib}
\usepackage{hhline}
\newcites{sec}{Primary Studies}
\usepackage{xcolor}
\usepackage{graphicx}
\usepackage{colortbl}
\usepackage{multirow}
\usepackage{tabularx}
\newcommand{\MyBox}[1]{\vspace{3mm}\noindent\framebox[\columnwidth][c]{\parbox[b]{0.95\columnwidth}{ #1 }}\vspace{3mm}}

\newcommand{\review}[1]{{#1}}

\begin{document}

\title{Women's Participation in Open Source Software: A Survey of the Literature}

  \author{Bianca Trinkenreich}
  \email{bianca_trinkenreich@nau.edu}
  \affiliation{%
  \institution{Northern Arizona University}
  \city{Flagstaff, AZ}
  \country{USA}}
  
 \author{Igor Wiese}
 \email{igor@utfpr.edu.br}
 \affiliation{%
 \institution{Universidade Tecnológica Federal do Paraná}
 \city{Campo Mourão, PR}
 \country{Brazil}}
  
 \author{Anita Sarma}
 \email{anita.sarma@oregonstate.edu}
 \affiliation{%
 \institution{Oregon State University}
 \city{Corvalis, OR}
 \country{USA}}
 
  \author{Marco Gerosa}
 \email{marco.gerosa@nau.edu}
 \affiliation{%
 \institution{Northern Arizona University}
 \city{Flagstaff, AZ}
 \country{USA}}
  
 \author{Igor Steinmacher}
 \email{igorfs@utfpr.edu.br}
 \affiliation{%
 \institution{Universidade Tecnológica Federal do Paraná}
 \city{Campo Mourão, PR}
 \country{Brazil}
  \institution{ and Northern Arizona University}
  \city{Flagstaff, AZ}
  \country{USA}
 }

\begin{abstract}

Women are underrepresented in Open Source Software (OSS) projects, as a result of which, not only do women lose career and skill development opportunities, but the projects themselves suffer from a lack of diversity of perspectives. Practitioners and researchers need to understand more about the phenomenon; however, studies about women in open source are spread across multiple fields, including information systems, software engineering, and social science. This paper systematically maps, aggregates, and synthesizes the state-of-the-art on women's participation in OSS. It focuses on women contributors' representation and demographics, how they contribute, their motivations and challenges, and strategies employed by communities to attract and retain women. We identified 51 articles (published between 2000 and 2021) that investigated women's participation in OSS. We found evidence in these papers about who are the women who contribute, what motivates them to contribute, what types of contributions they make, challenges they face, and strategies proposed to support their participation. According to these studies, only about 5\% of projects were reported to have women as core developers, and women authored less than 5\% of pull-requests, but had similar or even higher rates of pull request acceptances than men. 
Women make both code and non-code contributions and their motivations to contribute include, learning new skills, altruism, reciprocity, and kinship.
% thus it is important who may leave projects if they are not compensated for their contributions.
Challenges that women face in OSS are mainly social, including lack of peer parity and non-inclusive communication from a toxic culture. We found ten strategies reported in the literature, which we mapped to the reported challenges. Based on these results, we provide guidelines for future research and practice.
%%%%%%

\end{abstract}



%\ccsdesc[500]{Software and its engineering~Software evolution}
%\keywords{Open Source Software, FLOSS, Diversity, D\&I, Gender, Women.}

\maketitle

%\markboth{IEEE Transactions on Software Engineering}{}

%\IEEEpeerreviewmaketitle

%\thanks{B. Trinkenreich and M. Gerosa were with the School of Informatics, Computing & Cyber Systems, Northern Arizona University, Flagstaff, AZ, 86001 USA e-mail: bt473@nau.edu.}% <-this % stops a space
%\thanks{I. Steinmacher is with Federal University of Technology - Paraná - Brazil.}% <-this % stops a space
%\thanks{Manuscript received Month DD, YYYY; revised Month DD, YYYY.}}

\section{Introduction}

%{O}pen Source Software (OSS)development often involves a collaborative endeavor where expert developers, distributed around the globe create software solutions~\cite{forte2013defining,schrape2018open}. Several of these OSS projects count on a community of volunteers to shine and grow, and such a community needs newcomers for their sustenance and growth.
{T}he effects of gender\footnote{\label{foot1}In this study, we use the term ``gender'' as a socially constructed concept~\cite{butler1999gender}, where gender identification, display, and performance may or may not align with sex assigned at birth. To reflect this social concept of gender, in this paper we use the term ``women'' and ``men'' as a shorthand for people who self-identify as such.} diversity in Free and Open Source Software (OSS\footnote{in this study we use OSS, but other possible acronyms include FLOSS, F/OSS, or FOSS}) projects has gained increasing attention from practitioners and researchers~\cite{wurzelova2019characterizing,robson2018diversity,bosu2019diversity,vasilescu2015gender,catolino2019gender,fossatti2020gender}. Gender diversity has a positive effect on productivity as it brings different perspectives together, improving outcomes~\cite{vasilescu2015gender}, problem-solving capacity, and leading to a healthier work environment~\cite{earley2000creating}. A gender-diverse team has been shown to increase innovation and productivity in software engineering~\cite{ostergaard2011does,vasilescu2015gender,tourani2017code}.

%Diversity in software development teams can happen in many different ways, as gender, experience, culture, and specific technical knowledge~\cite{vasilescu2015gender}. 

Although OSS organizations have been taking actions to increase gender diversity and placing more women in leadership positions, the numbers are still low. Women represent only 5.2\% of the contributors in Apache Software Foundation~\cite{asf2016survey} and 9.9\% in Linux kernel~\cite{bitergia2016survey}, two of the largest and best-known OSS communities. 
Existing research suggests that gender bias and sexist behavior pervade OSS~\cite{nafus2012patches,gallego2015open}. 
%Women reported that feel frustrated when they are the only woman on a development team and when their input is not requested, despite their competence in certain areas of the project~\cite{lee2019floss}. 
While OSS projects idealize a strict meritocracy wherein quality speaks for itself~\cite{feller2000framework}, several biases undermine women, who feel their quality is not able to speak for itself and report experiencing ``impostor syndrome''~\cite{vasilescu2015gender}. Gender biases can represent a ``glass floor'' and a persistent barrier to entry~\cite{mendez2018open,padala2020gender}. To avoid negative gender-based biases, some women hide their gender by using neutral aliases in their profile~\cite{lee2019floss,vasilescu2015perceptions}. 

%Consistent with the general findings that women's participation in OSS remains very low and that OSS is considered hostile by some contributors, survey and anecdotal evidence have indicated that attracting and retaining women as contributors in OSS projects has been particularly challenging~\cite{ghosh2002free,weiss2005panel,wurzelova2019characterizing}. Previous studies went on a broader scope to cover user participation (of all genders) in a different domain (online communities)~\cite{malinen2015understanding}, or in the same population of interest (women) but on a broader scope both open source communities and software development projects in general~\cite{dias2019barriers}. 

Although many published scientific studies investigate women's participation in OSS, the findings are scattered across multiple fields and venues, and the results are disconnected from one another. 
%For instance, researchers have been trying to understand the low representation rate of women in OSS, as well as to learn more about their challenges and the strategies that can be adopted to attract and retain this underrepresented population. 
Summarizing and aggregating existing research can help communities fully comprehend women's participation in OSS to define proper mechanisms to increase the number of women contributors and produce new knowledge by connecting previously dissociated research.
Therefore, the purpose of this study is as follows: \textit{To review, summarize, and synthesize the current state of research on women's participation in Open Source Software.}

Through a systematic mapping of the literature, including database search, backward and forward snowballing, and input from prolific authors, we selected and retrieved information from 51 primary studies published between 2000 and 2021. Our contributions include characterizing who the women contributors to OSS are, what are their motivations to contribute, what types of contributions they make, what challenges they face while contributing, and what strategies can mitigate these challenges and support women's participation in OSS.
%the frequency of women's participation, the characteristics of the women who contribute to OSS, the ways women contribute and the acceptance rates of their contributions, their motivations to participate, the challenges they face as OSS contributors, and the strategies OSS communities can employ to help attract and retain women.

\section{Research Method}

We performed a systematic mapping of the literature (SML) because we expected the existing research to be fragmented and not follow common terminology or use established theoretical concepts. The systematic mapping of the literature is suitable in such circumstances because it maps the available evidence~\cite{petersen2008systematic,petersen2015guidelines}, categorizing and aggregating knowledge dispersed across disconnected studies. This section details our method, which is based on established guidelines~\cite{petersen2015guidelines,kitchenham2010systematic}.

\begin{figure*}[hbt]
\centering
\includegraphics[width=1\textwidth]{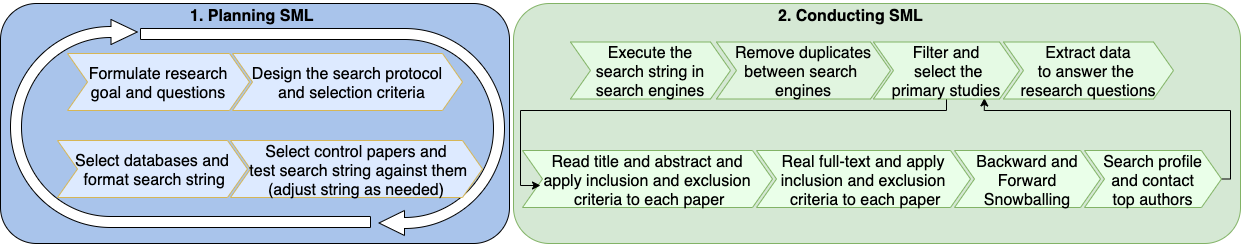}
\caption{Steps and phases of the Systematic Mapping of the Literature (SML), adapted from~\cite{kitchenham2007guidelines}}
\label{fig_method}
\end{figure*}

Fig~\ref{fig_method} depicts the steps and phases of our research method. Three researchers held four meetings to formulate the protocol. We piloted the search in Google Scholar and refined the search string, validating it with four control papers~\cite{robles2016women,vasilescu2015perceptions,terrell2017gender, canedo2020work}. We then executed the search string in the selected databases, conducted backward and forward snowballing, contacted the most prolific authors, applied the inclusion and exclusion criteria to filter the studies, and extracted data to answer the research question.

\subsection{Planning the Systematic Mapping of the Literature}

%In the planning phase, we formulated the search protocol, selection criteria, databases, and search string. We also defined control papers and piloted the study.

%\subsubsection{Research Goals}

To understand women's participation in OSS projects, we formulated the following research question: \textit{``How has the participation of women in OSS been} characterized in the literature?'' While reading the primary studies, we categorized the participation of women in OSS according to five perspectives that emerged from our analysis, as discussed in the results section.

\subsubsection{Search Protocol and Selection Criteria}
\label{selection_criteria}

We selected papers for our study based on the following Inclusion (IC) and Exclusion (EC) criteria. See supplemental material\footnote{https://figshare.com/s/6c54861e6829cb65f3cf} for additional details, including how the criteria were applied to each paper.

\begin{itemize}
    \item (+) IC1. The paper should characterize or be related to women's participation in Free/Open Source Software projects
    \item (-) EC1. The publication is just an abstract
    \item (-) EC2. The publication is not written in English
    \item (-) EC3. The publication is a copy or an older version of another article already considered
    \item (-) EC4. The publication is not a peer reviewed paper
    \item (-) EC5. The publication is a doctoral/master’s dissertation, research plan, short paper, or report
    \item (-) EC6. It was not possible to access the complete work.
\end{itemize}

\subsubsection{Databases and Search String}
\label{sec:db_selection}

We used a hybrid strategy~\cite{mourao2020performance} to collect primary studies, including database search (IEEE Xplore, Scopus, and ACM Digital Library)~\cite{petersen2015guidelines}, snowballing, and contact with the most prolific authors. These databases are suggested by Kitchenham and Charters~\cite{kitchenham2007guidelines} to conduct secondary studies in computer science and software engineering topics, and in our experience provide good coverage. Similar to related work~\cite{cosentino2017systematic,munir2016open,marques2014systematic,elberzhager2012reducing}, we limited the search to title, abstract, and keywords. Table~\ref{tab:search_string} provides the search string used for each database. 

\begin{table*}[htb]
\centering
%\scriptsize
\caption{Database search strings and results}
\label{tab:search_string}

\resizebox{\textwidth}{!}{\begin{tabular}{c|c|c}
\hline
\textbf{Database} &
  \textbf{Search String} &
  Results \\ \hline
Scopus &
  \begin{tabular}[c]{@{}c@{}}TITLE-ABS-KEY((woman OR women OR female OR gender OR  diversity OR "heterogeneous team") \\ AND ("open source" OR "open-source"  OR "free software" OR foss OR floss* OR oss))\end{tabular} &
  1067 \\ \hline
IEEE &
  \begin{tabular}[c]{@{}c@{}}("All Metadata":woman OR "All Metadata": women OR "All Metadata": diversity OR  \\ "All Metadata": female OR "All Metadata": "heterogeneous team" OR "All Metadata": gender) \\ AND ("All Metadata":"open source" OR "All Metadata":"open-source" \\ OR "All Metadata":"free software" OR "All Metadata": OSS \\ OR "All Metadata": FOSS OR "All Metadata": FLOSS*)\end{tabular} &
  350 \\ \hline
Compendex &
  \begin{tabular}[c]{@{}c@{}}(((((woman OR women OR female OR gender OR  diversity  OR  "heterogeneous team") \\ AND ("open source"  OR  "open-source"  OR  "free software" OR foss OR  floss* OR oss))))\end{tabular} &
  100 \\ \hline
ACM &
  \begin{tabular}[c]{@{}c@{}}Title:(((woman OR women OR female OR gender OR diversity OR  "heterogeneous team") \\ AND ("open source" OR "open-source" OR "free software" OR foss OR floss* OR oss))) \\ OR Abstract:(((woman OR women OR gender OR diversity) \\ AND ("open source" OR "open-source" OR "free software" OR foss OR floss* OR oss))) \\ OR Keyword:(((woman OR women OR gender OR diversity) \\ AND ("open source" OR "open-source" OR "free software" OR foss OR floss* OR oss)))\end{tabular} & 879 \\ \hline
\end{tabular}}
\end{table*}

The first part of the search string (before the AND) aims to limit the results to studies about women or the diversity and heterogeneity they can represent. The second part of the search string (after the AND) seeks to limit results to Free/Open Source Software. All searches were completed on November 15, 2020.

As we depicted in Table~\ref{tab:search_string}, to avoid missing studies that do not directly mention the terms ``women'' or ``female'' in the title, abstract, or keywords, we included in the search string more generic terms such as ``diversity'' and ``gender'', understanding that diversity can include other aspects (e.g. culture, age, tenure) and that gender can refer to genders beyond our focus on women.

%We did not restrict papers to those addressing women participation in OSS explicitly. Papers which did not address women in OSS but which did present a subgroup analysis per gender that had the information to extract for our research questions were selected.

%Aiming to maximize coverage, we first tried the search string using full-text. This produced an unmanageable number of papers (e.g. 11908 only in Scopus) with too many irrelevant results. We therefore revised the search string to cover only title, abstract, and key words.

\subsubsection{Control Papers and Piloting}
\label{sec:controlpapers}

Before searching the selected databases, we conducted exploratory searches using Google Scholar to refine the inclusion and exclusion criteria, research question, synonyms for the search string, and information extraction template. From that exploratory search and based on our experience, we selected four well-known, relevant studies to act as control papers~\cite{robles2016women,vasilescu2015perceptions,terrell2017gender, canedo2020work}. In our pilot studies to test the protocol, all four control papers were found.

\subsection{Conducting the Systematic Mapping of the Literature}

We executed the research protocol, summarized in Fig.\ref{fig_method}, which comprised the following six steps: execute the search string in selected databases, remove duplicates between databases using manual de-duplication, read and filter the papers applying IC and EC, perform the snowballing, and contact the most prolific authors to find additional studies.
Fig.~\ref{fig_method} shows the number of papers selected in each step of the selection process. Throughout the paper, we refer to the selected primary studies with ‘PS’ (e.g., (PS1)) to distinguish them from citations to other references (e.g., [1]).

\begin{figure}[hbt]
\centering
\includegraphics[width=0.5\textwidth]{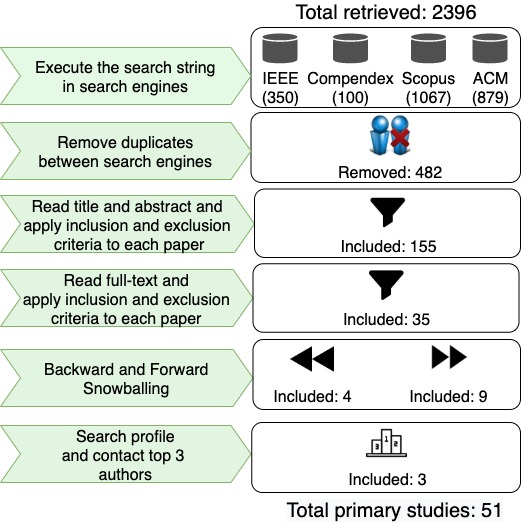}
\caption{Steps to filter and select the primary studies}
\label{fig_steps_select_primary_studies}
\end{figure}

\textit{Step 1: Search}. In November 2020, we performed the search on the selected search databases and found 2,396 papers.

\textit{Step 2: Remove Duplicates}. We used a spreadsheet to store and organize the retrieved papers, ordered the papers by title, and manually removed 482 duplicates.

\textit{Steps 3 and 4: Select Primary Studies}. Selection of papers followed the guidelines proposed by Kitchenham and Charters~\cite{kitchenham2007guidelines} and Petersen et al.~\cite{petersen2015guidelines} using the steps shown in Fig.\ref{fig_method}. 
%After the exhaustive search on the data sources using the search string, the papers retrieved were recorded and analyzed. We defined selection steps to help in filtering and including only the papers which provide direct evidence about the research questions and to reduce the likelihood of bias~\cite{kitchenham2007guidelines}. 
In the first step, we analyzed only the papers' title and abstract, consulting the entire text only when necessary to reach a confident judgment. In the second step, we filtered the papers considering the full-text reading. In both filters, papers were included to match the inclusion criteria or excluded if they fit any exclusion criteria.

\textit{Step 5: Backward and Forward Snowballing:} For Backward Snowballing, we conducted a single-step snowballing sampling, checking the primary studies references and looking for additional relevant papers. This step resulted in four additional primary studies. For Forward Snowballing, we also conducted a single-step sampling, checking what other papers cited the selected papers. This step resulted in nine additional primary studies.

\textit{Step 6: Contact prolific authors}. We reviewed the publication records and contacted the top 3 most prolific authors according to our primary studies, who are arguably specialists in the field. We searched for the Google Scholar profile and personal website of these authors and sent emails to them showing the list of primary papers and asking if they suggest others that we missed and should be included. This step resulted in three additional primary studies.

\begin{table*}[htb]
\centering
%\scriptsize
\caption{The categories of data to explain women's participation in OSS and the respective primary studies that provide data to answer them}
\label{tab:rqs_primary_studies}
\begin{tabularx}{\textwidth}{l|X}
\toprule

\textbf{Categories of Data - Women's Participation} & \textbf{Primary Studies} \\ \hline

1. Who are the women who contribute to OSS projects? & \cite{PS1,PS2,PS3,PS4,PS5,PS6,PS7,PS8,PS9,PS12,PS16,PS17,PS21,PS28,PS43,PS44,PS45,PS46,PS49,PS51}
\\ \hline

2. What motivates women to contribute to OSS projects? & \cite{PS5,PS16,PS18,PS19,PS50,PS51}
\\ \hline

%2. Demographics of women contributors to OSS & \cite{PS1,PS2,PS3,PS4,PS5,PS6,PS7,PS8,PS16,PS28,PS45,PS46}
%\\ \hline

3. What types of contributions do women make in\\ OSS projects? & \cite{PS2,PS4,PS5,PS6,PS9,PS10,PS11,PS12,PS13,PS14,PS15,PS45,PS46}
\\ \hline

4. What challenges do women face when contributing to OSS projects? & \cite{PS4,PS6,PS10,PS13,PS14,PS18,PS20,PS21,PS22,PS23,PS24,PS25,PS26,PS27,PS28}
\\ \hline

%5. What are the consequences of the challenges faced\\by women in OSS projects? & \cite{PS4,PS6,PS10,PS14}   
%\\ \hline

\begin{tabular}[c]{@{}l@{}}
5. What strategies were proposed to mitigate the challenges and to\\ support women's participation in OSS projects? \end{tabular} 
& \cite{PS3,PS6,PS8,PS9,PS12,PS16,PS23,PS24,PS25,PS26,PS27,PS28,PS29,PS30,PS31,PS32,PS33,PS34,PS35,PS36,PS37,PS38,PS39,PS40,PS41,PS42,PS47,PS48}
\\ \bottomrule
\end{tabularx}
\end{table*}

\subsection{Extracting and Analyzing Data}
\label{sec:extraction}

%We read the primary study, extracting the following information.

We extracted quantitative data from the papers to analyze what the literature had presented about women's representation in OSS projects and the rate of acceptance of women's contributions. 
We also qualitatively analyzed the papers and inductively applied open coding to create categories of data and organize what was reported by the literature about the demographics of women who contribute to OSS (Fig.~\ref{who_are_the_OSS_women}), their motivations (Fig.~\ref{motivations}), the types of contributions they make (Fig.~\ref{roles}), the challenges (Fig.~\ref{challenges}), and strategies (Fig.~\ref{strategies}). For types of contributions and motivations, the coding approach was informed by previous work \cite{von2012carrots,trinkenreich2020}. 
For challenges and strategies, we built post-formed codes and associated them with respective parts of the studies' text. Three of the authors conducted three card sorting sessions~\cite{Spencer2009} and discussed the codes and categorization until reaching consensus about the meaning of and relationships among the codes.

%The outcome was a set of higher-level categories as cataloged in our codebook \footnote{\label{foot1}\url{XXXXXX}}.

\subsection{Characterizing the Studies}

In this section, we present some characteristics of the primary studies (e.g., when they were published and who published them).

%\subsection{Where were the studies find?}

%Fig.~\ref{fig_venn} shows the distribution of the 35 papers selected from the search engines. The Compendex digital database search did not return any new publications, only duplicated ones.

%\begin{figure}[hbt]
%\centering
%\includegraphics[width=0.55\textwidth]{Figures/venn.jpeg}
%\caption{Distribution of the selected papers among the search engines}
%\label{fig_venn}
%\end{figure}

%\subsection{When were the studies published?}

\begin{figure}[hbt]
\centering
\includegraphics[width=0.66\textwidth]{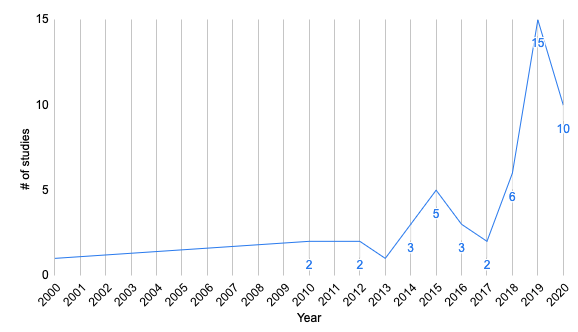}
\caption{Publications per year}
\label{fig_papers_year}
\end{figure}

Fig.~\ref{fig_papers_year} shows the number of primary studies published per year. The earliest paper was published in 2000. There were up to six papers per year until 2019, after which we saw a noticeable peak. We included one paper from 2021 through forward snowballing, but decided to not include it in Fig.~\ref{fig_papers_year} since it was outside the time period of our analysis.

%\subsection{Where were the studies published?}

%Fig.~\ref{fig_papers_venue} shows the most common publication outlets. The outlet with the most articles (7) is International Conference on Software Engineering (ICSE), followed by International Conference on Open Source Systems (4), Transactions on Software Engineering (3), and International Workshop on Gender Equality in Software Engineering (3).

%\begin{figure*}[hbt]
%\centering
%\includegraphics[width=0.70\textwidth]{Figures/papers_venue.png}
%\caption{Publication outlets (considering the outlets with more than one paper)}
%\label{fig_papers_venue}
%\end{figure*}

%\subsection{Who has published the studies?}
We ranked all 118 authors by the number of publications in the sample to investigate the most prolific authors. Among them, 95 authors had one paper; 17 authors had two papers; five authors had three papers; two authors had four papers; and one had 6 papers. The top authors were Alexander Serebrenik (10) and Bodgan Vasilescu (7).

\section{Results}

In this section, we present the results based on the perspectives that emerged from our analysis, namely the characterization of women's participation (Section~\ref{rq1}), motivation (Section~\ref{sec:rq5}), types of contributions (Section~\ref{sec:rq3}), challenges (Section~\ref{sec:rq6}), and the identification of strategies proposed in the literature to promote women's participation (Section~\ref{sec:rq7}).

\subsection{Who are the women who contribute to OSS projects?}
\label{rq1}

The literature presents different statistics for women's representation in OSS, which were measured in different ways, at different moments, and considering different sets of projects. We collected these results to provide a comprehensive view of the state of women's participation. Organizing data obtained from different sources may provide researchers with an expanded view of the phenomenon under study, prompting new insights and allowing for further discoveries~\cite{robles2014floss}. Additionally, analyzing women's representation at different time periods can help us understand the evolution of gender imbalance. Understanding the characteristics (e.g., education level, family status, diversification of projects, time to volunteer, tenure) can help communities and researchers refine their strategies to attract profiles of women who still do not participate and retain those who currently contribute to OSS.

Several primary studies quantified women's participation using various methods, including mining software repositories and mailing lists, surveys, and participation in mentorship programs. Table~\ref{tab:frequency_new} summarizes the frequency of women's participation reported by the primary studies.

\begin{table}[htb]
\centering
%\scriptsize
\caption{Frequency of women's participation in OSS as reported by primary studies. The gender was inferred using semi-automated tools in studies that used mining and self-reported by participants in surveys. We marked with * the studies conducted with data from a single project}.
\label{tab:frequency_new}

\begin{tabular}{r|r|l|c}
\hline
&\textbf{Data Source}&\textbf{Analyzed Population}                         & \textbf{\% Women} \\ \hline\hline
\multirow{26}{*}{\textbf{Mining}} 
%\multicolumn{2}{l}{\textbf{Mining}}                                           \\ \cline{3-4}\cline{3-4}
&\multirow{15}{*}{\textbf{Software Repositories}} &\multicolumn{2}{l}{\textbf{Hosting Platform Users}}                                            \\ \cline{3-4}
&&GitHub users (2015) \cite{PS17}            & 9\%                \\ \cline{3-4}
&&GitHub users (2019) \cite{PS46}            & 9.7\%               \\ \cline{3-4}
&&GitHub account owners (2018) \cite{PS16}   & 10.0\%                \\ \cline{3-4}
&&\multicolumn{2}{l}{\textbf{Change Requests' authors}}                            \\ \cline{3-4}
&&Authors of Pull-Requests on GitHub (2015) \cite{PS10}                 & 4.5\%    \\ \cline{3-4}
&&Authors of Pull-Requests on GitHub (2019) \cite{PS4} & 5.2\%               \\ \cline{3-4}
&&Authors of Pull-Requests on GitHub (2019) \cite{PS3} & 6.3\%               \\ \cline{3-4}
&&OpenStack code submissions on Gerrit (2017)  \cite{PS12} *                & 7.9\%  \\ \cline{3-4}

&&\multicolumn{2}{l}{\textbf{Commit authors}}                                   \\ \cline{3-4}
&&GitHub committers (2018) \cite{PS16}       & 5.9\%             \\ \cline{3-4}
&&Core Developers (2019) \cite{PS9}          & 4.3\%             \\ \cline{3-4}
&&Non-Casual Coders (2019) \cite{PS9}        & 6.7\%             \\ \cline{3-4}
&&OpenStack committers (2017) \cite{PS12} *   & 7.5\%             \\ \cline{3-4}
&&Authors of contributions to public code (2019) \cite{PS44}  & 10.0\%     \\ \hhline{~===}

&\multirow{3}{*}{\textbf{Mailing Lists}} &\multicolumn{2}{l}{\textbf{Participation in project-specific mailing lists}}\\   \cline{3-4}
&&Wordpress mailing lists (2014) \cite{PS49} * & 10.2\% \\ \cline{3-4}
&&Drupal mailing lists (2014) \cite{PS49} * & 8.5\%  \\ \hhline{~===}

&\multirow{8}{*}{\textbf{Google Summer of Code}} &\multicolumn{2}{l}{\textbf{Mentors}}                                          \\ \cline{3-4}
&&GSoC mentors (2016) \cite{PS43}            & 7.8\%             \\ \cline{3-4}
&&GSoC mentors (2017) \cite{PS43}            & 11.9\%            \\ \cline{3-4}
&&GSoC mentors (2018) \cite{PS43}            & 10.9\%            \\ \cline{3-4}
&&\multicolumn{2}{l}{\textbf{Students}}                                         \\ \cline{3-4}
&&GSoC students (2016) \cite{PS43}           & 11.9\%            \\ \cline{3-4}
&&GSoC students (2017) \cite{PS43}           & 13.4\%            \\ \cline{3-4}
&&GSoC students (2018) \cite{PS43}           & 14.2\%            \\ \cline{3-4}
\hline \hline
\multicolumn{2}{l|}{\multirow{3}{*}{\textbf{Survey}}}
%&\multicolumn{2}{l}{\textbf{Survey}}                                           \\ \cline{3-4}\cline{3-4}
&OSS contributors (2013) \cite{PS2,PS5}      & 10.4\%            \\ \cline{3-4}
\multicolumn{2}{l|}{}&OSS contributors (2019) \cite{PS21}         & 10.9\%            \\ \cline{3-4}
\multicolumn{2}{l|}{}&OSS contributors (2020) \cite{PS51}         & 7.6\%             \\ \hline
\end{tabular}
\end{table}

\textit{Participation of women in OSS found via mining software repositories:} From a dataset of 23,493 GitHub projects, Vasilescu et al.~\cite{PS17} used the genderComputer tool~\cite{PS49} (with 93\% of precision) to identify gender, based on personal names, and, if available, countries, of 873,392 GitHub contributors. They found 91\% men and 9\% women. From 5,250 OpenStack contributors, Izquierdo et al.~\cite{PS12} inferred gender using the genderize.io tool and found that 10\% are women. From a dataset of 8,338 GitHub projects, Prana et al.~\cite{PS16} found that the percentage of new GitHub accounts created by women remained around 10\% from 2014 to 2018. Bosu and Sultana~\cite{PS9} analyzed a dataset of 683,865 code review requests from 10 popular OSS projects. Authors inferred gender using the Gerrit-Miner tool~\cite{bosu2013impact} and found that women represent 6.70\% (out of 4,543) of non-casual developers (those who submitted at least five code changes) and only 4.27\% (out of 936) of core developers (those who are the top 10\% developers in terms of the number of code commits in a project). From a random sample of 300,000 GitHub users from a dataset with 16M users, Qiu et al.~\cite{PS46} inferred gender using genderComputer and NamSor tools and identified 9.7\% as women. Terrel et al.~\cite{PS4} analyzed a GitHub dataset with 4,037,953 profiles and identified the gender of 1,426,127 (35.3\%) users through their public Google+ profiles. Based on those profiles, the authors analyzed pull-request submission and acceptance by women and men. They found that 8,216 of the pull-requests were submitted by women (5.2\%) and 150,248 (94.8\%) by men. Imtiaz et al.~\cite{PS3} used the same GitHub dataset of \cite{PS4} and identified 529,253 men (93.7\%) and 35,676 women (6.3\%). Kofink~\cite{PS10} also analyzed a dataset of 1,811,631 pull requests and found that 4.5\% were submitted by women and 95.5\% by men. Zacchiroli~\cite{PS44} also analyzed the authors of contributions. With 1.6 billion commits from the combined projects of GitHub, GitLab, and other development forges (using the Software Heritage project\footnote{\label{Software Heritagefoot3}\url{https://www.softwareheritage.org/}}), corresponding to the development history of 120 million projects. Zacchiroli found that 33 million distinct people authored contributions over 50 years, and that there is a significant growth of active women as authors from around 4\% in 2005 to 10\% in 2019.
%and that in 2019 10\% of all contributions to public code were authored by women.

\textit{Distribution of women in OSS found by mining mailing lists:} Kuechler et al.~\cite{PS23} analyzed participation in eleven mailing lists of six projects (Buildroot, Busybox, Jaws, Parrot, uClibc, and Yum), which totaled 3,310 subscriptions. Authors found low participation by women: 8.27\% overall, 6.63\% of those who posted one message, 2.5\% of those who posted more than ten times, and 1.5\% of code reporters. Vasilescu et al.~\cite{PS49} also used the mailing lists of two projects (Drupal and Wordpress) to explore women's representation and found that women authored 9.81\% of the messages in Drupal and 7.81\% in Wordpress. In contrast, both men and women engage in OSS projects for statistically similar lengths of time.

\textit{Distribution of women who participate in mentorship programs:}
By analyzing the gender of Google Summer of Code participants from 2016 to 2018, Canedo et al.~\cite{PS43} found that while there is a minor variation across the years, the volume of women stayed close to 11.98\% of the total number of participants in the program. \footnote{\label{foot4}The Google Summer of Code (GSoC) is a 3-month OSS engagement program that offers stipends and mentorship to students willing to contribute to OSS projects~\cite{silva2017long,trainer2014community,silva2020google}.}

\textit{Distribution of women in OSS through surveys:} Mani and Mukherjee~\cite{PS2} and Robles et al.~\cite{PS5} analyzed the same OSS 2013 survey data \cite{arjona2014floss2013}. 2,183 OSS contributors answered the survey, 226 of whom identified as women (10.35\%). Lee and Carver~\cite{PS21} received 119 answers to their questionnaire, of which 10.92\% of respondents identified as women, while Gerosa et al.'s~\cite{PS51} questionnaire received 224 answers with 7.6\% who identified as women.

\begin{figure}[hbt]
\centering
\includegraphics[width=0.25\textwidth]{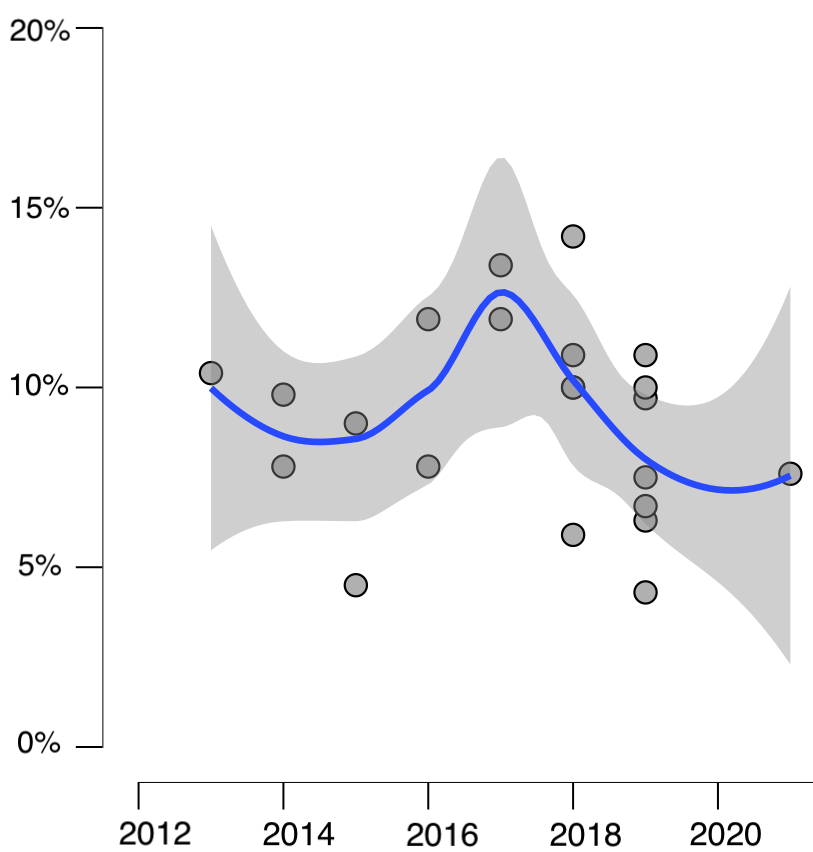}
\caption{The frequency of women's participation in OSS as reported by primary studies per year. The values comprise percentages of women measured by different mechanisms.}
\label{fig:rq1}
\end{figure}

In summary, the primary studies reported women's participation ratios ranging between 4\% to 14\% across different measurements and OSS communities. When we analyze the distribution over time based on when the primary studies were published (Fig \ref{fig:rq1}b), barring some fluctuation women's participation ratio stays stable at around 10\%. However, when taking a broader timeline view, Zacchiroli~\cite{PS44}, through his analysis of public code contributions over the last 50 years, found that women's contributions appear to be on the rise and are rising faster than those of male authors. This shows that while much still needs to be done, OSS projects are becoming more gender diverse, albeit slowly. 

%This study concludes that if the trend of the past 15 years is to continue, OSS authors and their contributions might soon reach a level of gender diversity comparable to other fields. 

%Considering the last 50 years, 7.5\% of all contributions to public code was authored by women. According to the author, 

%The ratio of women's commits has grown steadily over the past 50 years, reaching in 2019 for the first time 10\% of all contributions to public code.
 
%\input{Tables/rq1_frequency_of_women}

%\input{Tables/rq1_data}

%\MyBox{Women represent about 9.8\% of OSS contributors, considering different types of participation over the years.}
%The frequency of women's participation has been measured in different times, by either mining software repositories and mailing lists and using techniques (automatic and manual) to infer the gender of each contributor, by demographic questions in OSS surveys and registration in events and mentorship programs.}

%\subsection{Demographics of women contributors to OSS} 
%\label{rq2}

\begin{figure}[hbt]
\centering
\includegraphics[width=0.6\textwidth]{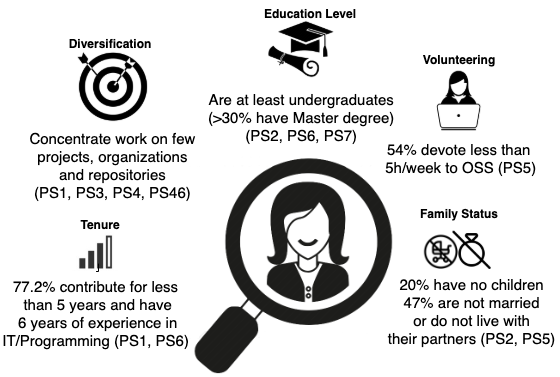}
\caption{The characteristics of women who contribute to OSS projects}
\label{who_are_the_OSS_women}
\end{figure}

To gain a deeper understanding of the demographics of women who contribute to OSS projects, we analyzed available data reported in the primary studies along the following criteria: education level, time dedicated to contributions, diversification of projects, family status, and tenure (Fig~\ref{who_are_the_OSS_women}). 

\textsc{Education Level.} Based on Stack Overflow's 2018 developer survey that included 43,000+ OSS contributors, Wurzelova et al.'s~\cite{PS7} study reported that 82.8\% of women who contribute to OSS are at least undergraduate students, compared to 76\% of the overall rate of contributors~\cite{overflow2018stack}. Mani and Mukherjee~\cite{PS2} found a similar rate (81.4\%) using data from the 2013 FLOSS Survey (226 women), while the corresponding figure for the whole dataset is 72\%. All of the 36 developers who identified as women on the survey from Canedo et al.~\cite{PS6} were at least undergraduate students. A substantial number of contributors from the three studies were post-graduates, who either achieved a Master's degree---11.6\%~\cite{PS6}, 27.6\%~\cite{PS7}--- or Ph.D.---4.2\%~\cite{PS7}, 10\%~\cite{PS2}, 22.9\%~\cite{PS6}.

\textsc{Volunteering.} Approximately half of the 226 women (53.59\%) who answered the 2013 FLOSS questionnaire devote less than five hours per week to OSS projects. Only 14.77\% of the women who answered this questionnaire dedicate more than 40 hours per week~\cite{PS5}, which can represent OSS as a full-time job. Although not mentioning the number of hours per week, according to the Powell et al.'s~\cite{PS28} results, 89\% of women said they contribute to OSS both at home and at work, which includes bringing their work home and contributing to OSS during their leisure time.

\textsc{Diversification of projects.} By mining software repositories, three studies~\cite{PS3,PS4,PS46} concluded that women concentrate their efforts on fewer projects than men. From a dataset of 152,534 pull-requests created by 20,926 women and 3,135,384 pull-requests created by 308,062 men, Imtiaz et al.~\cite{PS3} concluded that women's pull-requests are concentrated in fewer projects and fewer organizations than men's. Indeed, Qiu et al.~\cite{PS46} analyzed a balanced sample of the dataset, including 28,995 women and 29,096 men and also concluded that women tend to concentrate their contributions in fewer projects than men. From a dataset of 1,426,127 users whose gender could be identified, Terrell et al.~\cite{PS4} analyzed the acceptance of submitted pull-requests and concluded that women contribute to fewer projects than men. From another perspective, Vasilescu et al.~\cite{PS1} ran a survey answered by 199 women and 611 men, in line with the previous studies, and concluded that women own fewer public repositories than men. 

%\textsc{Age.} In different ways, four studies~\cite{PS1,PS5,PS6,PS8} show that most women who contribute to OSS are older than 20 and younger than 37 years old. According to the 226 women who answered the FLOSS 2013 questionnaire, the median age women join OSS is 26, the mean is 28, 1st Qu. is 22 and 3rd Qu. is 33, when many are already professionally active~\cite{PS5}. %Those women started in older ages than men~\cite{PS45}.
%Having a similar result, the median age of the 199 women who answered the Vasilescu et al.~\cite{PS1}'s questionnaire was 29, the mean is 31, 1st Qu. is 24 and 3rd Qu. is 37. More than half (52\%) of the 58 women who answered Singh's~\cite{PS8} questionnaire reported being 25-34 years old. Most of the 36 women who answered the Canedo et al.'s~\cite{PS6} questionnaire (80\%) were younger than 35 years old.

\textsc{Family Status.} Almost half of the 226 women who took part in the FLOSS 2013 survey are not married or did not live with their partners~\cite{PS2}. This rate was composed of 35\% single women, over 11\% not living with their partners, 3\% living with their partners, 3\% married, 0.1\% separated from their partners~\cite{PS2} and 20\% living with children~\cite{PS2,PS5}. 

%\textsc{Geolocation.} The majority of women who participated in the primary studies were from Brazil (44\%~\cite{PS1}), USA (37.1\%~\cite{PS1}---40\%~\cite{PS2}), and Germany (6\%~\cite{PS2}---9\%~\cite{PS16}---14.3\%~\cite{PS1}).
%Considering the overall contributors from all genders, the USA is top ranked (25\%~\cite{PS2}--32.6~\cite{PS1}), followed by Germany (6.5\%~\cite{PS1}--8\%~\cite{PS2}). Brazil had only 2.7\% of the overall contributors~\cite{PS1}, but 44\% of the overall women, showing that women stand out amongst contributors from this country.

%Although Brazil can apparently stand out in numbers, when Vasilescu et al.~\cite{PS1} aggregated countries in macro-regional level (Africa, Asia, Australia and New Zealand, Eastern and Southern Europe, Latin America, North America, Western and Northern Europe) and compared the distribution of respondents with a previous study of GITHUB users~\cite{takhteyev2010investigating}, and did not find difference. Vasilescu et al.~\cite{PS1} sent the invitation to answer the voluntary survey to a random sample of 4,500 GITHUB contributors, including individuals with a valid email addresses, different genders and who contribute to a different number of projects. The survey was about the perceptions of teams, diversity attributes among team members, and experiences with working in diverse teams. From the 816 received responses, 199 (24.4\%) respondents identified as a woman.

\textsc{Tenure.} Most (77.2\%) of the 199 women who answered Vasilescu et al.'s~\cite{PS1} questionnaire contributed to OSS projects for fewer than five years. Authors found that women have on average six years of experience in IT/programming, a significantly lower tenure than men with nine years of experience.

\MyBox{Women are underrepresented in central OSS roles, although they are better represented earlier in the joining process (e.g. in mentoring programs). Women's participation has increased in recent times, but those who contribute reported having a few hours per week to devote to OSS. Several studies found women's participation ranges from 4.3\% to 14.2\%.}
%, are at least undergraduate students, concentrate their work in a small number of repositories, projects, and organizations, and are mostly unmarried and with no children.

\subsection{What motivates women to contribute to OSS projects?}
\label{sec:rq5}

Research has shown that women are generally more motivated to use technology to accomplish a goal rather than for fun~\cite{burnett2010gender}. For the past 20 years, much academic work has theorized about and empirically examined OSS contributors' motivations. Retrieving and consolidating women's motivations from the existing studies is relevant to communities seeking to recruit and retain women. Proper management of motivation and satisfaction helps software organizations achieve higher productivity levels and avoid turnover, budget overflows, and delivery delays \cite{beecham2008motivation,francca2011motivation,DASILVA2012216}. 

\begin{figure}[hbt]
\centering
\includegraphics[width=0.8\textwidth]{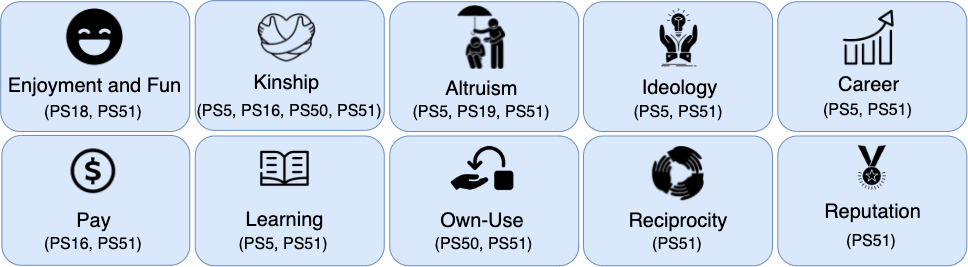}
\caption{Women's motivations to contribute to OSS. Based on~\cite{von2012carrots}}
\label{motivations}
\end{figure}

We consolidated the studies reporting women's motivation to participate in OSS projects, aggregating the results according to Von Krogh et al.'s categories~\cite{von2012carrots}. Von Krogh et al.~\cite{von2012carrots} surveyed the literature and identified ten categories of motivation, grouping them as intrinsic, internalized-extrinsic, and extrinsic. Intrinsic motivation (Enjoyment and Fun, Kinship, Ideology, and Altruism) moves a person to act for the fun or challenge entailed rather than in response to external pressures or rewards~\cite{ryan2000self}. In contrast, extrinsic motivations (Career and Pay) are based on outside incentives when people change their actions due to an external intervention~\cite{frey1997relationship}. Contributors can also internalize extrinsic motivators (Learning, Own-Use, Reciprocity, and Reputation) as self-regulating behavior, rather than as an external imposition~\cite{deci1987support,roberts2006understanding}. We summarize our findings organized in these higher level categories in Table~\ref{tab:motivations} and Figure~\ref{motivations}.

\begin{table*}[htb]
\centering
%\scriptsize
\caption{Women's motivations to participate in OSS projects}
\label{tab:motivations}

\begin{tabularx}{\textwidth}{c|c|X}
\toprule
\multicolumn{2}{c|}{\textbf{\begin{tabular}[c]{@{}c@{}}Categories   \cite{von2012carrots}\end{tabular}}} & \textbf{Women's motivations to participate in OSS}                                                                    \\ \hline
\multirow{8}{*}{Intrinsic}                                                                & Enjoyment and Fun               & Contribute when ``find it exciting" \cite{PS18}, have fun and feel stimulated by writing programs \cite{PS51}.
\\ \cline{2-3} 
 & Kinship           &  
 Work on a project with friends \cite{PS16} and participate in new forms of cooperation \cite{PS5}\cite{PS51}
\\ \cline{2-3} 
 & Ideology          & Believe that source code should be open and want to limit the power of proprietary software companies \cite{PS5} \cite{PS51}                      \\ \cline{2-3}
  & Altruism          & Share knowledge \cite{PS5}\cite{PS51}, improve other  developers' code \cite{PS5}, help people and practice philanthropy \cite{PS19}
  \\ \hline
\multirow{3}{*}{Extrinsic}                                                                & Career            & Improve career \cite{PS51} and have more job opportunities \cite{PS5}.                                             \\ \cline{2-3} 
& Pay & Being paid to contribute or selling products and services related to OSS \cite{PS16}\cite{PS51}.
\\ \hline

\multirow{4}{*}{\begin{tabular}[c]{@{}c@{}}Internalized\\ Extrinsic\end{tabular}}        & Learning          & Learn, develop and improve skills \cite{PS5}\cite{PS51}\\ \cline{2-3}

 & Own-Use          & Employment needs to use \cite{PS50} \cite{PS51} \\\cline{2-3}
  & Reciprocity          & \begin{tabular}[c]{@{}l@{}}Feel personal obligation because they use OSS \cite{PS51} \end{tabular} \\\cline{2-3}
  & Reputation          & \begin{tabular}[c]{@{}l@{}}Want to enhance my reputation \cite{PS51} \end{tabular}
\\ \bottomrule
\end{tabularx}
\end{table*}

%By interpreting the literature and our primary studies, we found both differences and similarities between general motivations to contribute to OSS~\cite{hertel2003motivation,harsworking,lakhani2003hackers,ghosh2002free,PS51} and women's motivations. %While \textsc{Enjoyment and Fun}, \textsc{Pay} and \textsc{Reciprocity} seem to have a different relevance for women; all other categories of motivations have similar importance for all genders.

Among all 10 categories, the literature showed that women see \textsc{Enjoyment and Fun}, \textsc{Reciprocity}, \textsc{Kinship}, and \textsc{Pay} differently from the motivation reported in the literature~\cite{hertel2003motivation,harsworking,lakhani2003hackers,ghosh2002free,PS51}---based on surveys answered predominantly by men. \textsc{Enjoyment and Fun} consistently has been reported as a top driver to contribute to OSS in motivation surveys~\cite{lakhani2003hackers,harsworking, ghosh2002free, hertel2003motivation,PS51}. However, according to the stratified analysis on Gerosa et al.'s~\cite{PS51} study, no women reported this motivation to join (and one--6\%--stayed because of this), while 7\% of the men who joined were motivated by fun and 20\% continued because of it. 
%Although in Balali et al.~\cite{PS18} we found one woman interviewee reporting that she contributes because she ``finds it exciting,'' 
%, but this motivation seems to not have the same relevance for women as to men.

The opposite trend is observed for \textsc{Reciprocity}, \textsc{Kinship}, and \textsc{Pay}. \textsc{Reciprocity} appeared as one of the top motivators for women in Gerosa et al.'s~\cite{PS51} work---39\% of the women reported this as motivation to continue (versus 15\% of the men). In general surveys~\cite{lakhani2003hackers,PS51}, reciprocity is not among the top motivators. Regarding \textsc{Kinship}, while 39\% of the 226 women who answered the FLOSS survey joined because of this motivation, 31\% continued because of it~\cite{PS5}. Most (64\%) of the 22 women who answered Prana et al.'s survey~\cite{PS16} select a project in which friends and colleagues also participate. As part of kinship, peer parity also plays a role in women's motivation~\cite{PS50}. David and Shapiro~\cite{PS50} found that social connections with other developers influence women's choices. However, no women from Gerosa et al.'s~\cite{PS51} study joined OSS because of kinship. Kinship is not top-ranked in general surveys~\cite{harsworking,lakhani2003hackers,ghosh2002free}, but this trend has been changing with the rise of social coding platforms~\cite{PS51}. Regarding \textsc{Pay}, Prana et al.~\cite{PS16} indicated that payment is a greater incentive for women than men (64\% vs. 35\%). This was echoed by women interviewees from Balali et al.~\cite{PS18}. However, the difference was not noticed in Gerosa et al.'s~\cite{PS51} work, in which men and women equally reported money as a reason to continue contributing (14\% and 11\%, respectively). 

%. One woman from Gerosa et al.'s~\cite{PS51} study reports ''Give back. A way to pay for my use of OSS''. In this study, 39\% of the women and 15\% of the men report that they contribute because of this motivation. Regardless of the importance for women, Lakhani et al's survey with all-genders contributors~\cite{lakhani2003hackers} report that reciprocity is not among the top motivators. 

%Pay

%Similarities
%According to Gerosa et al.'s~\cite{PS51} study, \textsc{Learning}, \textsc{Altruism}, and \textsc{Kinship}, are key motivations---on average 93\%, 85\%, and 80\% of the respondents (from all genders) contribute to OSS due to these motivations. We found those being similarly considered as women's motivations in primary studies.

For other motivations, we found that women follow a similar trend as reported in the general literature. \textsc{Learning}, for example, has been frequently reported as a key motivation to contribute to OSS \cite{lakhani2003hackers,hertel2003motivation,harsworking,ghosh2002free} and most (68\%) of the 226 women who took part in the FLOSS 2013 survey joined for \textsc{Learning}, and 65\% continue because of it~\cite{PS5}. 
%Two women (out of 18) from Gerosa et al.'s~\cite{PS51} study joined OSS to learn new skills, while three continued for that. 
The same happens for \textsc{Altruism}, which is a common motivator in OSS~\cite{PS51,ghosh2002free} and is relevant for women as well---37\% of the women who took the FLOSS 2013 survey reported that they joined to share knowledge~\cite{PS5} and 22\% from another work continued because of it~\cite{PS51}. 

%When analyzing 76 humanitarian OSS projects, Parra et al.~\cite{PS19} found that women are motivated to join this kind of OSS project to help people and having philanthropic objectives. Indeed, a high rate of the 226 women who took part in the FLOSS 2013 survey joined by \textsc{Altruism}, including 37\% to share knowledge and skills and 21\% to improve the OSS products of other developers. While no women from Gerosa et al.'s~\cite{PS51} study joined OSS by altruistic reasons, 22\% continued for that. 

%ideology

%Similarly, only two women (11\%) from this study joined OSS by ideology and one (6\%) continued for that. 

%Career

Similar rates were also found for \textsc{Career}, \textsc{Own-Use}, and \textsc{Reputation}. 
Regarding \textsc{Career}, only 4\% of the 226 women who took the FLOSS 2013 survey and none who took Gerosa et al.'s~\cite{PS51} survey reported joining to improve their careers. Similar rates were found for men (5\% and 8\%, respectively). For \textsc{Own-Use}, David and Shapiro~\cite{PS50} found that women are motivated by their employment-related needs and
one third (6 out of 18) of the women from Gerosa et al.'s~\cite{PS51} reported this motivation to join OSS. This is in line with previous research that shows that women are more motivated to use technology for what it enables them to accomplish~\cite{burnett2010gender}. \textsc{Reputation} was not a top motivation for contributors from all genders in the early 2000s~\cite{lakhani2003hackers,ghosh2002free} and still is not. Only 4\% of the men from from Gerosa et al.'s~\cite{PS51} joined because of reputation, while no women reported it.

%however, 26\% and 11\% reported that was their current motivation to stay. .
% , and 26\% continued for that~\cite{PS5}. Gerosa et al.'s~\cite{PS51} also found that few women start to improve their careers (0\%) 
%are also reported similarly by all-gender/men and women respondents in the motivation surveys. 

%Own-Use

%Reputation

Finally, an interesting case was \textsc{Ideology}, which was top-ranked in general surveys from the 2000s about motivation to join OSS~\cite{lakhani2003hackers,hertel2003motivation,harsworking,ghosh2002free}. Ideology is usually captured by motivations such as ``software should be free for all,'' ``free to modify and redistribute,'' or ``OSS should replace proprietary software.'' \cite{von2012carrots}. In the recent survey by Gerosa et al.~\cite{PS51}, this motivation has dropped some and was mentioned as a reason to join by only 11\% of the women and 11\% of the men. When considering only women, we can see that there was also a drop compared to the FLOSS survey, in which ideology was mentioned by 28\% of the women~\cite{PS5}. 

\MyBox{Reciprocity, Kinship and Pay are motivations especially relevant for women. On the other hand, enjoyment and fun motivate more men than women. Altruism, learning, and own-use motivate both men and women; career and reputation are low-ranked for both genders, and ideology was relevant in the past but has lost importance for both genders over the years.}

%\MyBox{Women are mainly motivated by learning and altruistic reasons, but payment also plays a role. Giving back and having more career opportunities retain women who are already contributors more than attract new women to join, being a more substantial reason to stay than to join OSS, while not getting paid can represent a motivation to leave.}

\subsection{What types of contributions do women make in OSS projects?}
\label{sec:rq3}

The OSS landscape has changed since the early 2000s to include the participation of ever more people and companies. Project-centric roles are becoming more established, and OSS projects increasingly include community-centric roles, which relate to areas beyond programming~\cite{trinkenreich2020}. Understanding how women contribute to OSS projects can help attract both code developers as well as those interested in being part of OSS via non code-centric roles.

\begin{figure}[hbt]
\centering
\includegraphics[width=0.6\textwidth]{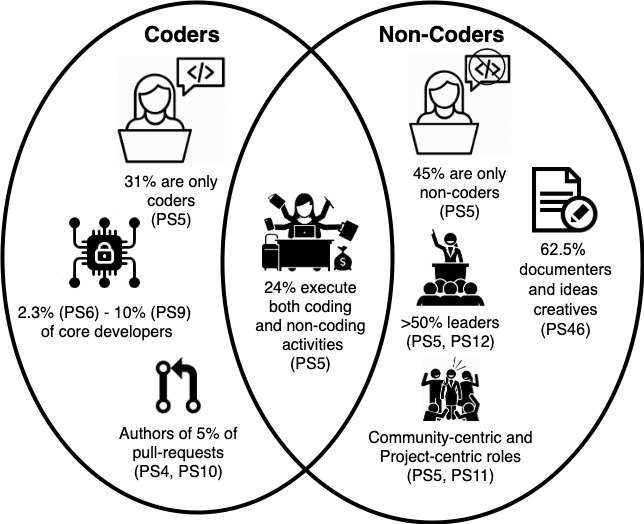}
\caption{Women make OSS contributions both as coders and as non-coders}
\label{roles}
\end{figure}

While the primary type of contribution in OSS projects is related to code development, the roles available in OSS projects go beyond the project-centric roles~\cite{trinkenreich2020}. They include many people who work ``behind the scenes'' to drive and sustain the community~\cite{PS11}. We categorized the activities reported by primary studies according to a framework of OSS roles from a previous study~\cite{trinkenreich2020} and with two perspectives: coders or non-coders, and project-centric or community-centric (Fig.\ref{roles}). Afterward, we present details about women's contributions as coders and non-coders reported by primary studies.

\begin{figure*}[hbt]
\centering
\includegraphics[width=0.8\textwidth]{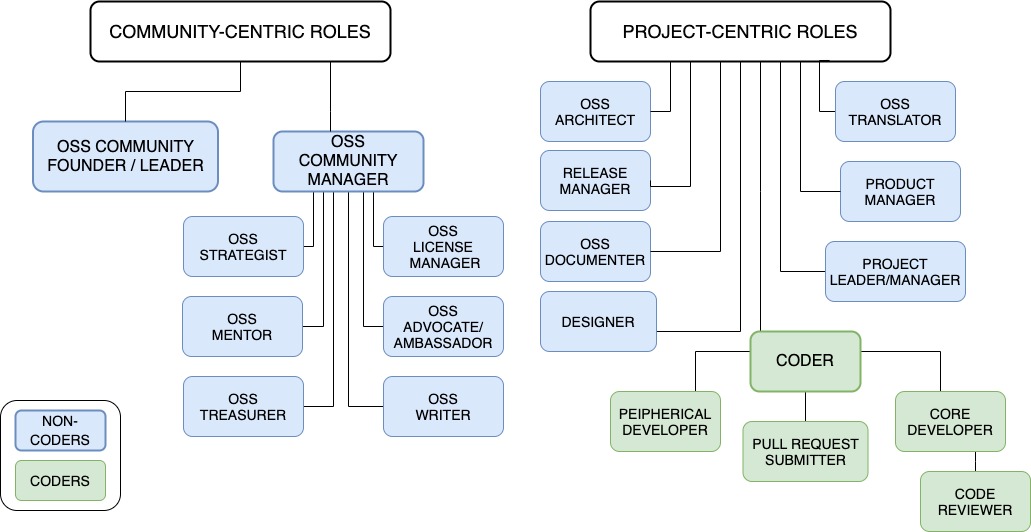}
\caption{Community-Centric and Project-Centric roles reported by primary studies as played by women who contribute to OSS. Roles can be played in parallel. Adapted from~\cite{trinkenreich2020}}
\label{centrics}
\end{figure*}

\subsubsection{Coders}

Only 31\% of the 226 women who answered the FLOSS survey~\cite{PS5} contribute to OSS projects as code developers, and 24\% perform coding in parallel with other roles.

\textsc{Core Developers.} The classic hierarchical model of coders in OSS development communities is described as a core-periphery structure, with a small number of core developers and a large set of peripheral developers~\cite{nakakoji2002evolution}. Core developers are code contributors involved with an OSS project for a relatively long time who make significant contributions to guide the project's development and evolution. Due to their relevant contributions and interactions, core developers often play leadership roles in OSS projects~\cite{ye2003toward}. Canedo et al.~\cite{PS6} found women as core developers in only 5.24\% of the 711 GitHub analyzed projects. Of all the core developers, only 2.3\% were women. From a dataset of 683,865 code review requests from ten popular OSS projects, Bosu and Sultana~\cite{PS9} found that women comprise a maximum of 10\% of core developers across all ten projects. Following a classification of commit types from Hattori and Lanza \cite{hattori2008nature}, Canedo et al.~\cite{PS6} concluded that women who are core developers contribute more with corrective and reengineering commits than forward engineering and management commits. Moreover, when describing the commits, women present a more detailed message explaining their contribution changes than men.
%which might indicate the gender bias symptom named proving-it-again, as women represent a group of people that does not align to the default stereotype (man) and demonstrate more evidence about their competence. 
\cite{PS15} evaluated the interactions in projects Angular.js, Moby, Rails, Tensorflow, Django, Elasticsearch and found that women who are core developers are more likely than men to interact with other contributors, evolve similarly to men within the project, and, though underrepresented, contribute to building sustainable social capital for OSS.

\begin{figure*}[hbt]
\centering
\includegraphics[width=0.9\textwidth]{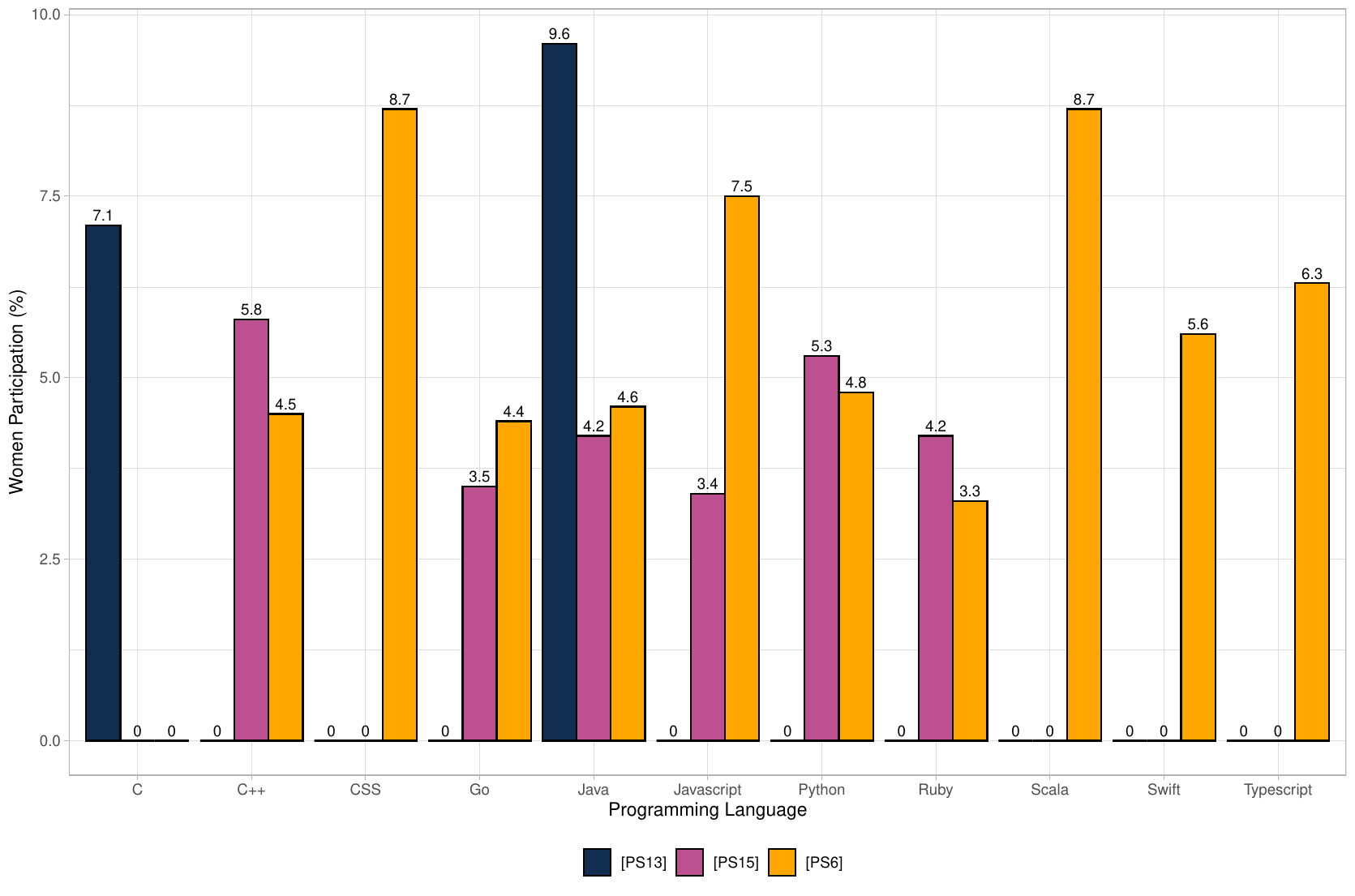}
\caption{Women's participation as core developers in OSS per programming language}
\label{programming_language}
\end{figure*}

\textsc{Preferred Technologies.} From the 711 projects analyzed by Canedo et al.~\cite{PS6}, women represented 8.8\% of the core developers in projects that are based on \textit{Scala} programming language, 8.7\% when based on \textit{CSS}, 6.3\% when based on TypeScript, 5.6\% when on based Swift, and only 1\% when the project is based on PHP and Shell programming languages. Still, from the same data, even projects written in TypeScript (10.93\%) have at least one woman as a core developer, and 2.17\% of projects using PHP have at least one woman as a core developer. Considering the data from the six projects analyzed by El Asri and Kerzazi~\cite{PS15} (Angular.js, Moby, Rails, Tensorflow, Django, Elasticsearch), women represented at least 4.8\% of core developers for the projects based on Python, at least 4.5\% for C++, and at least 4.2\% for Java. From the 158,464 of pull-requests for which Terrell et al.~\cite{PS4} could identify gender, women had a greater rate of accepted pull requests in Ruby, Python, and C++. We present the percentages of women as developers for each programming language as reported in the primary studies in Fig~\ref{programming_language}.

\textsc{Code reviewers.} Peer code review is a practice in software engineering. A code developer submits the code produced to another person (peer) to evaluate and find possible errors before merging the code to the project codebase~\cite{bacchelli2013expectations}. According to the GitHub dataset of six projects evaluated by Paul et al.~\cite{PS13}, women are more likely to write reviews expressing sentiments in the text to another woman than to a man during code reviews. Huang et al.~\cite{PS14} used medical imaging and eye-tracking to evaluate the visual and cognitive processes and patterns of neural activation followed by reviewers while performing code reviews. Authors found that women spent significantly more time analyzing pull request messages and author pictures (regardless of their identity) than the code itself when performing code reviews. 
%Huang et al.~\cite{PS14} also analyzed the behavior of code reviewers. Using the visual and cognitive processes and patterns of neural activation followed by reviewers while performing code reviews, authors concluded that women spent

\subsubsection{Non-Coders}

Almost half (45\%) of the 226 women who answered the FLOSS 2013 survey take part in non-coding activities, and 24\% perform code-related activities in parallel with other roles~\cite{PS5}.

\textsc{OSS Community Manager.} After analyzing the career pathways followed by 17 contributors, Trinkenreich et al.~\cite{PS11} presented a set of community-centric roles, including community founders and managers, strategists, mentors, writers, license managers, treasures, and advocates. The contributors who play these roles are usually ``hidden figures,'' who are not visible when analyzing the data from projects' repositories or coding platform websites. 11 (out of the 12) women interviewed in this study play community-centric roles.

\textsc{Project Leader/Manager.} More than half (51.49\%) of the 226 women who answered the FLOSS 2013 survey participate in community leader, coordinator, or administrator roles, while only 5\% of those women coordinate more than three projects~\cite{PS5}. Women in OpenStack who play leadership roles represent 7\% of the project committee members, 8\% of the project team leaders, 9\% of the project board directors, 7\% of the technical committee, 8\% of the working group leaders, and 23\% of project ambassadors~\cite{PS12}.

\MyBox{Women are more present in community-centric than in project-centric roles; almost half make non-code contributions. However, they often play both coding and non-coding activities in parallel. Very few projects have at least one woman as a core developer.} 
%When reviewing code, women also read pull-request messages and analyze the author's pictures.

\subsection{What challenges do women face when contributing to OSS projects?}
\label{sec:rq6}

Previous work investigated challenges faced by OSS contributors who are mentors and newcomers~\cite{steinmacher2015social,balali2018newcomers,mendez2018open,steinmacher2016overcoming}. Some of them report gender bias as a challenge~\cite{balali2018newcomers}, and others report barriers faced by women~\cite{dias2019barriers}. 
Anecdotes about gender bias appear across the literature~\cite{terrell2017gender,kofink2015contributions}, and women have been reported to feel that such biases are to blame for their contributions' comparatively low acceptance rate~\cite{canedo2020work}. 
  %Indeed, women’s acceptance rates are higher only when they are not identifiable as women~\cite{terrell2017gender}. Implicit bias are cultural, rather than as a result of innate differences~\cite{kofink2015contributions}. 
We aggregated the scientific evidence about challenges and gender bias. Understanding the nuances of women's challenges to contribute to OSS can help communities plan strategies to mitigate these challenges and thereby attract and retain more women.

Women mainly face socio-cultural challenges when contributing to OSS~\cite{PS21}, which can also influence their decision to leave an OSS project~\cite{PS13}. From the 37 women who answered Powell et al.'s~\cite{PS28} survey, 50\% indicated they had witnessed gender-based discrimination within the OSS community either online, in meetings, or in class, and 50\% said they had experienced harassment online or offline. Gender-related incidents can be so severe that they motivate women to leave an OSS project~\cite{PS1}. Indeed, in another survey~\cite{PS21} women reported that they drop out when the OSS project does not prioritize diversity. Leaving an OSS project is a decision that impacts more women than men---according to Qiu et al.~\cite{PS46}, women are 27\% more likely to disengage from GitHub. Understanding the reasons behind the decisions to step out of a project can help create strategies to increase retention in OSS. Kuechler et al.~\cite{PS23} suggest that women drop out because the OSS project is not aligned to their motivations or due to the unappealing and hostile social dynamics in projects. 

\begin{figure}[hbt]
\centering
\includegraphics[width=0.7\textwidth]{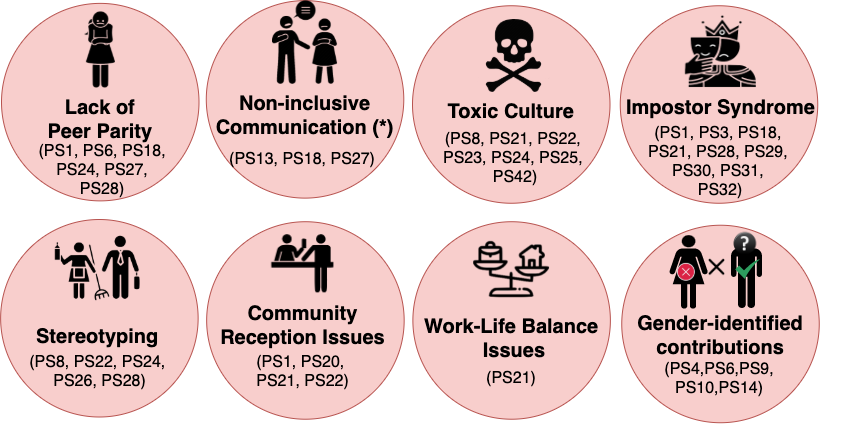}
\caption{Challenges faced by women when contributing to OSS}
\label{challenges}
\end{figure}

We summarize the challenges found in the literature, which were all socio-cultural, in Table~\ref{tab:challenges} and Fig.~\ref{challenges} and mark with an asterisk (*) the ones reported as a challenge that ultimately can cause women to leave OSS. Next, we present and explain each of them.

\begin{table*}[htb]
\centering
%\scriptsize
\caption{Challenges faced by women when contributing to OSS projects}
\label{tab:challenges}
\begin{tabularx}{\textwidth}{c|X}
\toprule

Challenge                                                                 & \multicolumn{1}{c}{Description}                                                                                                                                                                                                                \\ \hline
\multirow{3}{*}{\shortstack[c]{Lack of Peer Parity}}                             & Women feel alienated~\cite{PS28}, frustrated~\cite{PS1}, invisible~\cite{PS24} and less comfortable without other women around~\cite{PS18}, specially in medium-size projects where all contributors~\cite{PS27}.  \\ \hline

\multirow{2}{*}{\shortstack[c]{Non-Inclusive\\ Communication}}  & The expletives often used in the mailing lists~\cite{PS13}, documentation~\cite{PS18} and code reviews~\cite{PS27} are insulting to women and can cause them to leave an OSS project~\cite{PS13}.                                                                                                                      \\ \hline
\multirow{3}{*}{\shortstack[c]{Toxic  Culture}}                             & Incidents of symbolic violence, harassment~\cite{PS24}  and sexism against women~\cite{PS22} bring hostility~\cite{PS25} and can hinder their access to the community~\cite{PS21}, as they may already be hesitant about how they will be received~\cite{PS23}.                                    \\ \hline
\multirow{3}{*}{\shortstack[c]{Impostor Syndrome}}                                                & Even being competent~\cite{PS32} and knowing the importance of confidence~\cite{PS28}, women face a lack of self-efficacy~\cite{PS18}~\cite{PS21,PS29,PS30,PS31}, are more restrained~\cite{PS3} and reluctant to publicly display their work~\cite{PS1} than men in general \\ \hline

\multirow{3}{*}{\shortstack[c]{Community\\ Reception Issues}}      & Finding a mentor is a hard task~\cite{PS21}, as men consider as a dating opportunity~\cite{PS22}, which leads many women to hide their gender~\cite{PS1} and feel restrained when being repealed by communities for not being ready to provide contributions soon~\cite{PS20}.                                     \\ \hline
\multirow{2}{*}{\shortstack[c]{Stereotyping}}                             & Women are being boxed into specific roles~\cite{PS24}~\cite{PS26}, sometimes treated by men as if they were their mother for who they ask how to dress and behave~\cite{PS22}                                                                                                                                                                  \\ \hline
\multirow{1}{*}{\shortstack[c]{Work-Life Balance Issues}}    & Lack of time due to family responsibilities~\cite{PS21}                                                                                        \\ \hline
\multirow{3}{*}{\shortstack[c]{Gender-Identified\\ Contributions}} & Women have lower pull-request acceptance rate when they explicitly identify themselves as women~\cite{PS4}, wait longer for reviews and feedback~\cite{PS9}, and reported to perceive gender  bias against their contributions~\cite{PS6}.                                                              \\ 
\bottomrule
\end{tabularx}
\end{table*}

%Women who answered the Lee and Carver~\cite{PS21}'s survey reported they stand out an OSS project when they have the impression that the project do not care about diversity.

\textsc{Lack of Peer Parity.} Most women (72\%) feel outnumbered and 24\% feel alienated~\cite{PS28}. Women reported to feel more comfortable and accepted by their same gender counterparts~\cite{PS18} and feel frustrated when there is no peer parity~\cite{PS1}. This problem worsens in medium-size projects in which all contributors are men, as they may form a clique that a woman contributor could have difficultly breaking into~\cite{PS27}. \textit{``It is not so common to find many girls in technical teams''}~\cite{PS6}, and in the face of this lack of parity women reported feeling invisible in larger men-dominated groups~\cite{PS24}.

\textsc{Non-inclusive communication.} Discriminatory expletives, swear words, and negative critiques often used in code reviews and mailing lists may be insulting to women. The negative workplace experience of encountering words that are demeaning to them in the mailing lists can cause women to leave OSS projects~\cite{PS13}. Awkward communication styles~\cite{PS18} in acrimonious talk about which code piece should be incorporated~\cite{PS22}, and terms usually associated with men (e.g., ``guys'') can demotivate women~\cite{PS27}.

\textsc{Toxic Culture.} Incidents of symbolic violence and harassment against women can hinder their access to the community, such as when men decide to `hire that one because she is hot'~\cite{PS24}. Geek-saturated communities like Slashdot are often unwelcoming and hostile environments~\cite{PS25}. Additionally, women are sexualized in OSS~\cite{PS8}, facing judgment, abuse, hostility, and discrimination~\cite{PS42}. While hurtful and offensive talk is openly addressed to them, women are obliged to remind men not to ``stare and point'' at them~\cite{PS22}. From the 13 women who answered a questionnaire sent to contributors of 15 OSS projects, 38\% had suffered some incidents of sexism, including sexist statements or assumptions, being ignored, insinuations that they had it easy because they were women, and being simultaneously held to higher standards than men and underestimated~\cite{PS21}. Moreover, the authors found that women had trouble being taken seriously and needed to prove themselves (prove-it-again~\cite{PS3}). According to~\cite{PS23}, public flaming and aggression can be enough to distort women's participation, as they may already be hesitant about how they will be received.

\textsc{Impostor Syndrome.} Women tend to be risk-averse \cite{dohmen2011individual} and have low computer self-efficacy \cite{burnett2010gender,burnett2011gender,cazan2016computer,hartzel2003self,huffman2013using,singh2013role}, which can affect their behavior with technology, causing women to be less confident in their ability to complete tasks and blame themselves if there is a problem~\cite{PS29,PS30,PS31}. Women find it challenging to directly translate competence to confidence without social attraction (being liked by the other community members in terms of having a rapid increase of followers). Consequently, initiating a pull-request to a new repository can be problematic due to women's competence-confidence gap~\cite{PS32}. Even understanding that confidence is an essential factor when entering OSS~\cite{PS28}, women face a lack of self-efficacy~\cite{PS21,PS18}. \textit{``Despite having proved [their] competency in certain areas of the code/project, [their] opinion is rarely or never asked for''} (quotation from Vasilescu et al.~\cite{PS1}). Still, Imtiaz et al.~\cite{PS3} found that women tend to be more restrained than men in general. Despite being knowledgeable and professionally well-settled, women may be more reluctant to publicly display their work ~\cite{PS1}.

\textsc{Community Reception Issues.} Women reported to feel restrained when communities nullify them when they do not have enough skills to provide contributions on their first day~\cite{PS20}. When trying to find a mentor, upon discovering their mentee's gender, men mentors can treat the relationship as a dating opportunity~\cite{PS22}. This makes finding a mentor an arduous task, which includes attracting attention and breaking into a close-knit OSS community~\cite{PS21}. Many women use fake GitHub accounts and hide their gender, \textit{``so that people would assume [they] were male''}~\cite{PS1}.

\textsc{Stereotyping.} Pre-existing stereotypes~\cite{PS8,PS28}, gender roles, and ``macho'' attitudes can cause gender inequalities in OSS communities~\cite{PS24}. Women are boxed into specializations despite their manifest protest against it, as the legal case against the front-end/back-end distinction has shown~\cite{PS26}. Additionally, men often treat women as if they were their mothers, asking for advice about how to dress and behave and then refusing to enter into a technical dialogue thereafter~\cite{PS22}.

\textsc{Work-life Balance Issues.} Women that participated in Lee and Carver's~\cite{PS21} study reported a lack of time and family responsibilities. Only women from this study, and not from other studies, reported family responsibilities as a challenge.

\textsc{Gender-Biased Peer-Review}
%As a consequence of the challenging environment, women often decide to hide their gender when contributing to OSS. Some study to The literature have been analyzing the bias against women's contributions by the merge rate of pull-requests as a measurement of women's contributions' acceptance.
%For code contributions, we used the merge rate of pull-requests as a measurement of women's contributions' acceptance. 
%Kofink~\cite{PS10} analyzed 1,811,631 pull-requests from 1,049,345 different users and found that women's pull-requests account for only 4.5\% of the total. Terrell et al.~\cite{PS4} presented a similar rate: women submitted 5\% of the pull-requests (from 158,464 in which authors could identify gender) and tend to have their pull-requests accepted at a slightly higher rate (78.7\%) than men (74.6\%), regardless of experience level. Authors found that, while less experienced developers making their initial pull-requests do get rejected more often, women generally still maintain a higher rate of acceptance throughout. Even submitting fewer pull-requests, according to Kofink~\cite{PS10}, women have 64.2\% of merge acceptance rate, nearly equivalent to men's (63.9\%), and continue to have high acceptance rates as they gain experience. According to Huang et al.~\cite{PS14}, humans (non-bots) are 4.7\% more likely to accept pull-requests submitted by women than men. 
Even in a population of core developers, one-third of the 36 women who answered Canedo et al.'s~\cite{PS6} survey reported they believe that reviewers had not accepted at least one of their contributions due to gender bias. Moreover, 11.4\% of the women participants recognize gender bias while someone assesses their contributions. Although women can have merge acceptance rates nearly equivalent or slightly higher than men~\cite{PS4,PS10}, according to Terrell et al.~\cite{PS4}, there is a bias against women's contributions when the gender is identified. The authors found that women have a 12\% lower acceptance rate when they explicitly identify themselves as women, comparable to 3.8\% for men who disclose their gender. However, authors found that women tend to have their pull-requests accepted at a slightly higher rate (78.7\%) than men (74.6\%) when not identifying the gender, regardless of experience level. While less experienced developers making their initial pull-requests do get rejected more often, women generally still maintain a higher rate of acceptance throughout. Bosu and Sultana~\cite{PS9}'s study corroborated this bias that women face by using three additional metrics (first feedback interval, review interval, and code churn per comment). The study analyzed ten projects and found explicit biases against women in three of the analyzed projects (Android, Chromium OS, and LibreOffice). Women had lower code acceptance rates than men, and had to wait longer to receive initial feedback for their code changes and to complete code reviews. The code submitted by women also had lower churn per comment in both Android and Chromium OS. The study showed that Android and LibreOffice stand out in prominent gender biases, where women had 10\% lower acceptance rates then men, and review intervals that last three times longer than men. On the other hand, three other projects indicated biases favoring women (oVirt, Qt, and Typo3).

%Robles et al.~\cite{PS5} showed that women participate less if they have children.

%\input{Tables/challenges}

\MyBox{Women face many socio-cultural challenges that include: missing having other women around (lack of peer parity), encountering offensive language in mailing lists (non-inclusive communication), suffering symbolic violence and harassment (toxic culture), avoiding initiating a pull-request due to lack of confidence (impostor syndrome), facing challenges finding mentors (community reception issues), being boxed into specific roles (stereotyping), sharing time between work and family (work-life balance issues), and biases or lower acceptance rates of contributions when they explicitly identify themselves as women (gender-biased peer review).}

\subsection{What strategies were proposed to support women's participation in OSS projects?}
\label{sec:rq7}

Recommendations of strategies to improve diversity are scattered and rarely widely adopted. OSS communities need a concise view of the different types of actions to select the ones that are viable and appropriate for their needs and for the challenges they face. Strategies include actionable mechanisms that OSS communities can take and combine to create a more inclusive environment for women. 

\begin{figure}[hbt]
\centering
\includegraphics[width=1\textwidth]{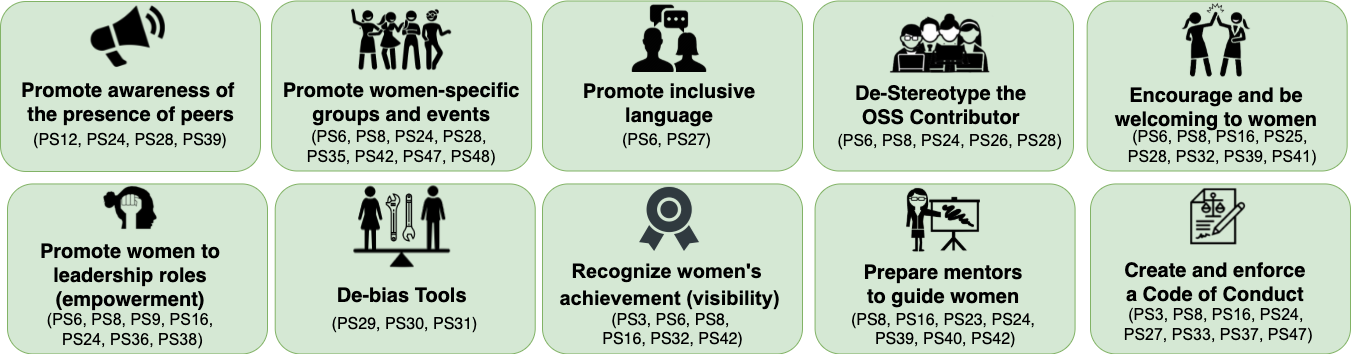}
\caption{Strategies employed by OSS communities and projects to increase women's participation in OSS}
\label{strategies}
\end{figure}

We summarize the proposed strategies in Fig.~\ref{strategies} and Table~\ref{tab:strategies}. Next, we present and explain each of them. All strategies had been mentioned as a way to mitigate at least one challenge presented in Sect.~\ref{sec:rq6}. However, there was no strategy reported to mitigate the challenge \textsc{Work-Life Balance Issues} nor \textsc{Gender-biased Peer Review}.

\begin{table*}[htb]
\centering
%\scriptsize
\caption{Strategies to increase women's participation in OSS projects}
\label{tab:strategies}

\begin{tabularx}{\textwidth}{c|X}
\toprule

Strategy
&
\multicolumn{1}{c}{Description}                                                 \\ \hline
\multirow{2}{*}{\shortstack[c]{Promote awareness of\\the presence of peers}}
&
Reduce the lack of peer parity by having more women involved \cite{PS24} \cite{PS28} and propagating their participation \cite{PS12} \cite{PS39}.        \\ \hline

\multirow{4}{*}{\shortstack[c]{Promote women-specific\\groups and events}}
&
Awake technological vocations with schoolgirls' OSS events \cite{PS24} and code development training \cite{PS35}, create forums and   women-only spaces \cite{PS6} \cite{PS8} \cite{PS24} \cite{PS42} \cite{PS28}, and facilitate women's registrations and participation in events \cite{PS47}.
 \\ \hline

\multirow{2}{*}{\shortstack[c]{Promote inclusive \\language}}
&
Avoid  gender pronouns \cite{PS6} that assume that people are [all] the same gender \cite{PS27}.
 \\ \hline

\multirow{2}{*}{\shortstack[c]{De-stereotype the \\OSS contributor}}
&
Fix and do not stereotype activities on gender \cite{PS8} \cite{PS28}, as code development for men \cite{PS6} \cite{PS26} and community   management for women \cite{PS24}. 
 \\ \hline
\multirow{3}{*}{\shortstack[c]{Encourage and be \\ welcoming to women}}
&
Have women encouraging other women \cite{PS16} \cite{PS25} to make code \cite{PS28} and also non-code contributions \cite{PS32}, increasing  their confidence \cite{PS6}, while making the community friendly and supportive \cite{PS8} \cite{PS39} \cite{PS41}. 
 \\ \hline
\multirow{2}{*}{\shortstack[c]{Promote women to\\leadership roles   (empowerment)}}
&  
Train women in leadership skills \cite{PS9} \cite{PS38}, promote them to senior, decision-making positions, \cite{PS6} \cite{PS8} \cite{PS16} \cite{PS24} and mediating \cite{PS36} roles.
 \\ \hline
\multirow{3}{*}{\shortstack[c]{De-bias tools}}
& 
Use the Gendermag technique to find bias in tools, and use recommendations to adjust the software, making it more inclusive to  the women’s cognitive style \cite{PS29} \cite{PS30} \cite{PS31}. 
  \\ \hline
\multirow{4}{*}{\shortstack[c]{Recognize women's\\achievement (visibility)}} &

Showcase the success of women \cite{PS16} \cite{PS42}, celebrate their achievements \cite{PS3}, recognize their efforts \cite{PS32} and give credits   when they deserve it \cite{PS8}, using either the communication channels \cite{PS6} and events in which they can be speakers \cite{PS24}. 
 \\ \hline
\multirow{3}{*}{\shortstack[c]{Prepare Mentors to\\Guide Women}}           &
Train mentors to guide women \cite{PS8} \cite{PS39} \cite{PS40} \cite{PS42}, bring cultural proximity between mentors and mentees \cite{PS16}, making  sure novices find the help and support they need \cite{PS23} \cite{PS24}. 
 \\ \hline
\multirow{5}{*}{\shortstack[c]{Create and enforce\\a Code of Conduct}}
&
Develop a code of conduct \cite{PS3} \cite{PS16} \cite{PS27} \cite{PS47} as the collective norms on (un)acceptable behavior at all interactions,  explicit the   prohibition of harassment \cite{PS21} \cite{PS24} \cite{PS37}, and that violations have consequences \cite{PS8} \cite{PS33}, while having mechanisms to invigilate the use and to apply the punishments accordingly if necessary.
\\\bottomrule
\end{tabularx}
\end{table*}

\textsc{Promote awareness of the presence of peers.} Approximately half (54\%) of the women respondents of Powell et al.'s~\cite{PS28} study said they would be more inclined to participate in OSS if there were more women involved. Promoting awareness about the rate of women can help to attract more women, minimizing the feeling of alienation~\cite{PS28}. This awareness should include a measurement of women's participation and the type of contributions~\cite{PS12,PS39}. According to participants of Calvo's~\cite{PS24} study, the communities can create parallel spaces in which the proportion of women is above 50\% so as to create more diverse environments under the values of mediation and care.

\textsc{Promote women-specific groups and events.} The community managers interviewed by Calvo~\cite{PS24} mentioned that they promote schoolgirls' events to inspire vocations and empower girls who may opt for an OSS career. %Helping women to join the code development teams can be done by investing in women-specific educational and professional training efforts and outreaching for women interested in code development~\cite{PS35}. 
This strategy should promote activities exclusively for women in those spaces, highlighting their presence, as women tend to be invisible in larger groups in which men are the majority~\cite{PS24}. For those women already interested in OSS, promoting women-only groups, spaces, and events~\cite{PS6,PS8,PS24,PS42,PS28,PS47} fosters discussions and supports networking and empowerment~\cite{PS42}. Moreover, it provides a safe space for expressing feelings and opinions~\cite{PS24} and revealing their identities~\cite{PS47}. Although effective, Singh and Brandon~\cite{PS47} found that only 3\% of the 350 projects they analyzed have women-specific spaces---including websites, IRC Channels, dedicated blogs, collection/list of resources, dedicated Facebook pages, and local meet-ups. 
%(as men sometimes monopolize debates). 
%According to Singh and Brandon's~\cite{PS48} study in mailing lists, women use the forums for networking and feel safe and comfortable to reveal their identities. 
%Besides the women spaces, Singh and Brandon~\cite{PS47} pose that communities should encourage the participation of women in events, either prioritizing their registrations or proposing them as participants. 

\textsc{Promote inclusive language.} The insights provided by Qiu et al.~\cite{PS27} include avoiding gender pronouns that assume that people are [all] one gender or one demographic. For example, using ‘guys’ is common, and it can imply that contributors are men~\cite{PS6}.

\textsc{De-stereotype the OSS contributor.} The women interviewed in Singh's~\cite {PS8} study recommended to %not isolate women; include them in activities, conferences, and
``leave the stereotypes out the door.'' The frustration caused by stereotypes was expressed by one of the women surveyed by Canedo et al.~\cite{PS6}: ``Stop treating women developers as 'women developers' and start treating them as developers.'' Powell et al.~\cite{PS28} suggests showing less discrimination and more inclusion to tone down the male-dominated atmosphere and promote participation. Calvo~\cite{PS24} and Vedres and Vasarhelyi~\cite{PS26} suggest avoiding the \textit{feminization} of specific assignments, like those relating to community building tasks; OSS communities should re-classify types of work that used to be packaged in masculine-feminine stereotyped specialties.% Following the idea of de-stereotyping, authors suggest caution when planning the coding schools for women to avoid training them in things that already have several women working (such as Ruby).

\textsc{Encourage and welcome women.} Singh's~\cite{PS8} findings show that being less judgmental and appreciating diverse teams is essential to supporting and encouraging women. Indeed, as Beach~\cite{PS41} discusses, people need to feel supported, accepted, and encouraged. This encouragement may even come from other women~\cite{PS16}. Powell et al.~\cite{PS28} suggested starting by encouraging small steps as an incentive to women to submit bug reports and share their input, increasing their self-confidence~\cite{PS25}, which was echoed by two participants of Canedo et al.'s~\cite{PS6} study, ``the solution is to build confidence'' and ``not to fear when contributing.''
%According to Wang et al.~\cite{PS32}, when contributing to an OSS project, even with non-code contributions, women tend to increase self-confidence. Parker~\cite{PS25} presents some some specific initiatives to encourage women to become contributors, as: 
Encouragement is also the goal of some initiatives presented by Parker's~\cite{PS25} FLOSSpols\footnote{\label{foot6}\url{www.flosspols.org}}, which offers recommendations on how to solve the gender gap; these initiatives include WOWEM, a gender equity and OSS research and education project; and LinuxChix, a community for supporting women in Linux. There is no value in encouraging women to be there if the environment is hostile. To welcome women, one of the community managers who participated in Barcomb et al.'s~\cite{PS39} study recommends making the community friendlier in general.

\textsc{Promote women to leadership roles (empowerment).} A way to empower women is to have them in senior roles~\cite{PS6}, project governance~\cite{PS38}, and where appropriate~\cite{PS8}, as mentioned by one woman participant of Prana et al.'s~\cite{PS16} study: ``More women reviewers. More women are acting directly on the governance of large OSS projects.'' Some community managers indicated that their communities created decision-making positions and ensured that women led public activities~\cite{PS24}. 
Catolino et al.~\cite{PS36} 
%study analyzed the community smells of a dataset including 20 projects that have only men contributors and other 20 projects having also at least one woman. Authors showed that, even when outnumbered, women can act as mediators against the proliferation of community smells. Authors
suggested that a way to avoid the proliferation of community smells is to involve women in positions where they can mediate discussions and improve the communication of sub-communities. According to Singh~\cite{PS8}, promoting women to positions of authority shows the project respects their contributions. 
%Finally, Bosu and Sultana~\cite{PS9} and Qiu et al.~\cite{PS38} recommend to prepare and mentor them to develop their leadership skills, to help women to become leaders.

\textsc{De-bias tools.} Most (73\%) of the barriers that affect software professionals demonstrate some form of gender bias~\cite{PS29}. Indeed, bias in tools and infrastructure can hinder women newcomers from joining OSS ~\cite{PS29}. One way to de-bias infrastructure and tools is by applying the GenderMag technique. GenderMag uses personas and a specialized Cognitive Walkthrough (CW) to systematically evaluate software and make them more inclusive of women's cognitive styles~\cite{PS29,PS30,PS31}. This technique's precision was proven by a study that showed that the GenderMag technique helped to identify 81\% of the issues~\cite{burnett2016gendermag}.

\textsc{Recognize women's achievement (visibility).} When recognizing women's achievements, the community provides the social attraction that women seek to overcome their competence-confidence gap~\cite{PS32}. Communities can show recognition by increasing the visibility of women~\cite{PS16}, listing them as great contributors whenever they deserve it~\cite{PS42}, and \textit{publicly} celebrating their achievements in blogs, project homepages, and social media~\cite{PS3,PS16}. Another way to increase visibility is to organize events where speakers are women~\cite{PS24,PS47}. These simple actions inspire more women to participate~\cite{PS16,PS24}.

%Singh~\cite{PS42} suggested to showcase the success of women by providing listing and contact information for top contributors. After concluding that women tend to moderate their behavior for the sake of worrying about other people's expectations, Imtiaz et al.~\cite{PS3}, also suggested to form a group to celebrate each others’ accomplishments, since women are less to boast about their own achievements, a traditionally masculine behavior. Not only women, but also men who answered the Prana et al.'s~\cite{PS16} questionnaire mentioned that having an explicit visibility of other women's achievement can inspire women to participate. Canedo et al.~\cite{PS6} suggest the visibility can be created by publishing women' work in communication channels (blogs, forums, public speaking), while Singh~\cite{PS8} weighs to give credit to women ``when they deserve it''. Another way to operate visibility that was mentioned by a Calvo's~\cite{PS24} interviewee is to organize events where all the speakers are women. Although they are small activities, this idea gave visibility to women.

\textsc{Prepare Mentors to Guide Women.} As mentioned before, OSS projects are usually men-dominated, which may deter women~\cite{PS22,turkle2005second}. Mentorship can help women newcomers find the assistance and support they need~\cite{PS23, PS24, PS39}. One way to do so is to have women mentor other women~\cite{PS8, PS42}.
%., if possible having mentors and mentees relatively close in region or even close in cultural dimension~\cite{PS16}. 
Singh~\cite{PS8} highlights that when mentoring women it is necessary to guide them on different aspects. While men need to change their behavior and projects need to implement systemic changes, Singh~\cite{PS8} posits that women also need to be trained to ignore disruptions and not be easily bothered by criticism or insults; the mentor needs to be extra supportive, friendly, respectful, and encouraging~\cite{PS8}. %Mentors can develop specific tactics to increase perceived usefulness among women regarding the acceptance of OSS solutions so they start using it~\cite{PS40}. %dvise them to be persistent, don’t give up, and keep trying, find a mentor in the community, choose a small project with a community known to support women, collaborate with other women, be unfazed, and develop a thick skin and ignore disruptions, be supportive, friendly, respectful and encouraging~\cite{PS8}. Mentors can develop specific tactics to increase perceived usefulness among women regarding the acceptance of OSS solutions so they start using it~\cite{PS40}.

\textsc{Create and enforce a Code of Conduct.} Developing a code of conduct for the community~\cite{PS3,PS8,PS16,PS24,PS27,PS33,PS37,PS47} helps to mitigate Tightrope effects\footnote{\label{foot5} The term Tightrope is usually associated with the circus, where a performer walks on a stretched rope. Following the analogy, the term can refer to the narrow band of socially acceptable behavior for someone, in our case, women~\cite{PS3}.} by assisting communities in articulating acceptable behaviors for all members~\cite{PS8}. The code of conduct comprises the collective norms of a community, as mantras that shape the culture of collaboration~\cite{PS37} as well as the community’s expectations and values to create a friendly and inclusive community~\cite{PS8,PS33}. %In practice, the code of conduct includes the set of rules on acceptable and unacceptable conduct at all interactions (messages, meetings, conferences), and emphasizes the respect for minority groups, and the explicit prohibition of harassment~\cite{PS24,PS37}. The code of conduct can serve as mechanisms to advertise the intention to have more women through explicit statements that they welcome women participants and have zero tolerance for behavior contrary to this position. 
While having a code of conduct will not prevent sexism, it indicates to any men who display sexist behaviors that such actions will not be tolerated in the project~\cite{PS21}. However, according to Robson's study~\cite{PS34}, just creating a code of conduct will not increase women's participation. The author showed that projects that introduced a code of conduct in their history saw women's participation increase almost at the same rate of projects without a code of conduct. Projects increased from 2.37\% to 3.81\% after a code's introduction. Projects without codes of conduct, comparing gender diversity within similar periods yields, had an increase from 4.10\% to 5.53\%. The average increases were 1.44\%, and 1.43\%, respectively, which means creating the code of conduct did not help increase women's participation. Robson~\cite{PS34} posits that the code of conduct needs to be enforced among the project members. Communities should put mechanisms in place to implement the code and show that violations have consequences~\cite{PS8,PS33}.

\begin{figure*}[hbt]
\centering
\includegraphics[width=1\textwidth]{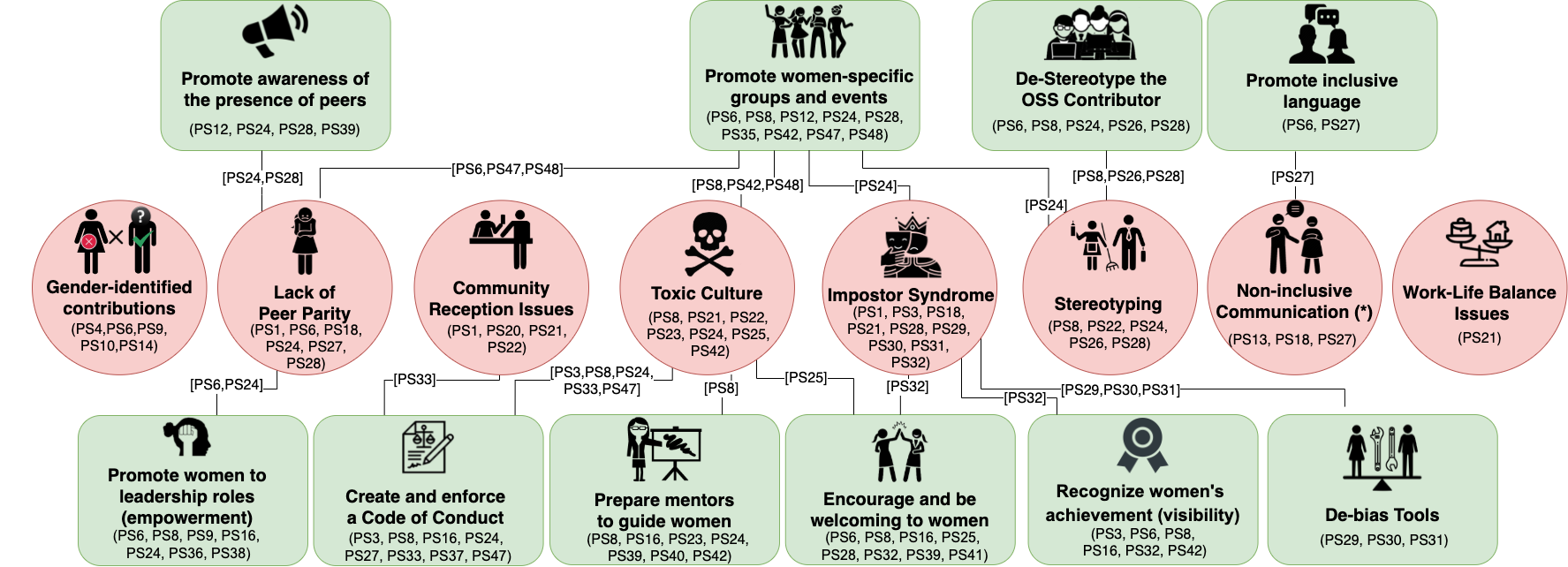}
\caption{The strategies that were mentioned by primary studies to mitigate challenges faced by women in OSS (circles represent challenges and rectangles represent strategies)}
\label{strategies_challenges}
\end{figure*}

\MyBox{The strategies that OSS communities can employ to increase women's participation include providing awareness and statistics about women contributors, inclusive language, women-specific groups and events, de-stereotyping the OSS contributor, encouraging and welcoming women, having women in leadership, de-biasing tools, recognizing women's achievement, preparing mentors to guide women, and creating and enforcing a code of conduct.}

\section{Discussion}
\label{sec:discussion}

In this section, we discuss our results, strengths of evidences of the reported strategies to increase women's participation, gaps in literature, and opportunities for future research.

\subsection{Challenges in improving diversity}

\textbf{The leaky OSS pipeline.} Section~\ref{rq1} reports that there is a large gender disparity in OSS contributors (Kofink~\cite{PS10} found less than 5\% of pull-requests authors were women). However, the gender disparity is less pronounced in the initial stages (e.g., as students of Google Summer of Code). There is attrition of women contributors as they move through the different stages of the ``joining script''~\cite{von2003community}---where people start outside the project as readers and passive users; then move to the project periphery as bug fixers, bug reporters, peripheral developers, and active developers (developers without commit rights); and finally enter the project as core members or project leaders~\cite{nakakoji2002evolution}---the leaky pipe phenomenon. This attrition can be a consequence of the several socio-cultural challenges faced by women during the process. As we presented in Section~\ref{sec:rq6}, women face gender bias in communication, acceptance in the community, and lower contribution acceptance rates when they explicitly identify themselves as women.
While mentorship events enhance (women) participants' sense of competence and increase the chances of future contributions' values~\cite{silva2020theory}, these programs alone are insufficient as women do not stay long enough to become project leaders.
The majority of the challenges that women face or the reasons for women leaving OSS (see Section \ref{sec:rq6}) are socio-cultural in nature and not related to technical skills. Indeed, as we presented in Section \ref{sec:rq6}, women have similar or even higher rates of merge acceptance than men. Besides the appropriate technical skills, Bosu and Sultana's study~\cite{PS9} suggested that women are as productive as their colleagues who are men in an inclusive OSS project, with some women developers showing more productivity than the average men developers. Therefore, strategies that help create an inclusive environment geared towards retaining and mentoring women are needed to fix the leaky pipeline.

\textbf{Impostor syndrome amplified by toxic culture.}
Impostor syndrome is a psychological concept about a pattern of behavior wherein people (even those with adequate external evidence of success) doubt their abilities and experience a persistent fear of being exposed as a fraud~\cite{mullangi2019imposter}. As we presented in Section~\ref{sec:rq6}, the literature shows that when women join OSS, the rate of acceptance of their contributions is similar to---if not higher than---men's. The \textsc{Impostor Syndrome} in OSS can be amplified by the hostile and \textsc{Toxic Culture} that pervades communities, which are two of the challenges reported by primary studies presented in Section~\ref{sec:rq6}. The impostor syndrome disproportionately affects women and other minority groups, who often lack sufficient role models of success~\cite{mullangi2019imposter}. Combining actions that include training for allies, plus both reactive and proactive mechanisms to enforce the code of conduct among the project members~\cite{PS34} can help mitigate the toxic culture, as we present in Section \ref{sec:discussion:strategies}.
 
%{AS: what to do: ally training...to give more people the vocabulary to do something if they see toxicity? Have COCs that are enforceable}

\textbf{Impact of gender stereotyping.} Research shows conflicting findings about the effects of diversity on team performance. Although low diversity can enhance mutual trust and effectiveness, the demographic similarity may also lead to stereotyping, cliquishness, and conflict \cite{de1994expectation,molleman2006impact}. Regardless of the cause, according to a survey of 5,500 GitHub users~\cite{zlotnick}, women more often than men encounter language or content that makes them feel stereotyped. Stereotypes manifest the common expectations about members of certain social groups. Both descriptive (how women are) and prescriptive (how women should be) gender stereotypes and the expectations they produce can compromise a woman's career progress~\cite{heilman2001description, heilman2012gender}. Even before starting a career, stereotype threats represent one of the significant barriers to underrepresented groups engaging in Computer Science education. Implicit stereotypes about gender and STEM have profound effects on girls' and women's interest, confidence, and persistence in STEM education and careers~\cite{dasgupta2014girls,dasgupta2011ingroup}.

A fear of gender stereotyping can lead women to hide their gender~\cite{PS1} and create pseudonyms to avoid judgment~\cite{PS21}. This behavior was also observed by Ford et al.~\cite{ford2019beyond} in online communities, where participants use a ``gender neutral alias for websites like technical communities, because [they] get better help when asking questions or answering them.'' One of the strategies presented in Section~\ref{sec:rq7} is \textsc{De-stereotyping the OSS contributor}, which differs from using a neutral username to hide gender. In fact, Canedo et al.'s~\cite{PS6} study showed that users who do not reveal their gender suffer an even more severe disadvantage in survival probability. Although it prevents discrimination by categorical gender, avoiding gender identity can also lead to a lack of trust and exclusion from projects and ultimately cause a higher exit rate of such users. Even when stereotyping is minimal, it can still make a difference. People's attitudes, beliefs, and behavior are often shaped by factors that lie outside their awareness~\cite{banaji1998consciousness, greenwald1995implicit}. Considering that even minimal social cues may activate negative stereotypes early in informational processing~\cite{wu2020influence}, \textsc{De-stereotyping the OSS contributor} is crucial for women to start seeing themselves playing the role of developer and not just men. This strategy is aligned with the suggestion of women interviewed in the Blincoe et al.'s study~\cite{blincoe2019perceptions}, who considered that changing the typical image of software engineers as IT geek men is a way to reduce the gender gap.

\subsection{Considering women contributor's motivations} 

\textbf{Motivation to join vs. project culture.} Since women have social motivations (e.g. \textsc{Kinship}) as we presented in Section~\ref{sec:rq5}, and the reported challenges are also social, as we presented in Section~\ref{sec:rq6}, there is a conflict between their expectations and reality, which can explain why women are not joining or staying in OSS projects. When women join an OSS project expecting to find other women~\cite{PS50} and friendly colleagues~\cite{PS16}, but instead find \textsc{Lack of Peer Parity} and face a \textsc{Toxic Culture}, this directly conflicts with their motivations. One place where toxicity can manifest in communications is via comments in code reviews and in mailing lists. Code reviewers may need education support to articulate their review comments in a way that builds relationships~\cite{bosu2016process}. One strategy could be providing \textit{review templates} that help developers in using inclusive words and employing empathy. The lack of peer parity can be alleviated by attracting more women, which can be accomplished by recognizing different types of contributions~\cite{PS11}, for example recognizing contributors who participate by answering questions and discussing issues~\cite{trinkenreich2021,ducheneaut2005socialization}. Moreover, communities can foster peer communication through women-only (and their ally) groups and events, such as R-Ladies and other safe spaces~\cite{PS6,PS42,PS48}.
%AS:SO WHAT CAN WE DO ABOUT IT :D? review templates...the R-Ladies/women safe spaces work by Vandana?/ change culture from code hackers to acknowledge other forms of contribution, such as mentoring}

\textbf{OSS as a career pathway.} As Section~\ref{sec:rq5} presents, only 4.07\% of the 226 surveyed women from a FLOSS 2013 study joined to increase their job opportunities. After becoming contributors, this motivation increased almost six times (going to 25.79\%)~\cite{PS5}. We argue that this represents the ``shifting belief'' that women have in OSS toward building a career, which increases only after overcoming the barriers to join and become contributors. The multiple roles presented in Section~\ref{sec:rq3} are both related to the technical (project-centric) and non-technical (community-centric) side of the projects. Therefore, an awareness of the different roles and career pathways that exist in OSS can attract women with diverse backgrounds and expertise to OSS by showing them the multitude of trajectories to success~\cite{PS11}. Programs like Google Summer of Code, as well as other OSS-academic liaisons, can improve awareness of career opportunities provided by OSS. Further, given that 54\% of the women who contribute to OSS devote less than 5 hours per week, this means that the majority do not making a living from OSS. The fact that \textsc{work-life balance issues} are a challenge that women face (see Section \ref{sec:rq6}) and \textsc{being paid to contribute} is a relevant motivation (see Section \ref{sec:rq5}), OSS projects could offer part-time jobs to attract women who are not yet participating.

\subsection{Choosing appropriate strategies to mitigate challenges}
\label{sec:discussion:strategies}

\subsubsection{The strengths of evidence for recommended strategies} 

The literature has suggested several strategies to mitigate challenges faced by women and increase their participation in OSS (Section \ref{sec:rq7}). However, these strategies have not yet been evaluated in the field. Therefore, to help communities in selecting or prioritizing strategies to plan interventions, we analyzed the strengths of evidence of the publications that reported them.
To grade the strength of evidence, we followed a process similar to~\citet{dybaa2008strength}, using the GRADE (Grading Recommendations Assessment, Development and Evaluation) working group definitions~\cite{grade2004grading}, which grades the overall strength of the evidence as high, moderate, low, or very low. The GRADE approach is widely used by secondary studies in software engineering to evaluate the strength of evidence found in the primary studies~\cite{dybaa2008empirical,alves2010requirements,alexandre2014state,silva2015using,guinea2016systematic,gurbuz2018model}.

%Evaluating the strength of evidence includes the results' validity of primary studies. The 
GRADE suggests categorizing the strength of evidence of the studies by combining four key elements: study design, study quality, consistency, and directness. Randomized experiments are evaluated at the top
of the hierarchy (high) and evidence from observational studies and expert opinion are at the bottom of the hierarchy (low). In addition, the initial overall grade can be increased by the evidence base or decreased depending on the seriousness of the study's limitations. The strength of evidence is defined by GRADE as the extent to which one can be confident that an estimate of effect or association is correct. While inconsistent, imprecise, sparse, or biased reported evidence may decrease the grade of an experiment, strong evidence of association from two or more high-quality studies may increase the grade of an observational study~\cite{grade2004grading,dybaa2008strength}.

\textsc{Study Design:} All the primary studies were case studies that reported strategies recommended by participants and none evaluated interventions through controlled experiments. Thus, as per GRADE, the evidence from these observational studies is classified as ``low.'' Consequently, our initial categorization of the total evidence based on (primary) study design is also low. 

\textsc{Consistency} relates to the similarity of estimates of effect across studies~\cite{grade2004grading}. Since the outcomes of primary studies were not presented in a comparable way and reporting protocols vary according to the study, evaluating the synthesis of quantitative results could be misleading. However, having several studies grouped into each strategy (see Fig.\ref{strategies} reflects positively on the consistency of the studies and we conclude that, in general, strategies are consistent.

Regarding \textsc{Directness}---or the extent to which the people, interventions, and outcome measures are similar to those of interest~\cite{grade2004grading}---we found that most studies had another goal besides presenting the strategy(ies). However, little effort has been made to present evidence about implementing a particular strategy to increase women's participation in OSS projects. Most of the strategies were suggested as part of interviews or discussed by authors based on other findings. The only three strategies that formed part of the main goal of the papers were: use of the GenderMag technique (\cite{PS29,PS30,PS31}, enforcing a code of conduct (\cite{PS34}), and providing women-only spaces~\cite{PS8}. From those three, only Robson's study \cite{PS34} evaluated women's participation with and without the reported strategy. However, this study showed that many confounding factors may have influenced the results. Our judgment was thus that there are major uncertainties about the directness of the included studies.

Finally, although the overall \textsc{quality} of the studies was high, with well-described methods and data collection, the threats to validity were presented but not always addressed. Therefore, in the absence of any evaluation or superficial assessment of the effectiveness of the proposed strategies, the quality of the results (effectiveness of the strategy) is graded as very low. Combining the four components of study design, consistency, directness, and study quality, we found that the strength of the evidence about the strategies to increase women's participation is low and uncertain. Therefore, while many primary studies have shown the different challenges that women face, there remains a gap in research about the success of the proposed strategies to mitigate these challenges.

\subsubsection{Combining synergistic strategies.} 
One option for communities looking to improve diversity is to combine synergistic strategies. This might be especially useful since the literature has identified many different strategies, which have low strength of evidence. OSS communities can start by implementing simple, but structured actions combining ideas from more than one strategy. For example, by publishing success stories of women in the media, OSS communities can \textsc{promote awareness of presence of peers} to attract more women and also \textsc{recognize women's achievement (visibility)} to retain women who are already contributors. Considering that this media exposure can include women's posts and pictures, this action also helps with the strategy of \textsc{de-stereotype the OSS contributor}, which has been associated with images of men in technical textbooks~\cite{makarova2015trapped,lee2018gender} and search results~\cite{kay2015unequal}. Another action that can use more than one strategy is to create a women-only forum, which is part of the strategy to \textsc{promote women-specific groups and events}. When moderating and analyzing the messages from women to implement feasible changes to problems that are being actively discussed, this action also acts to \textsc{encourage and be welcoming to women} by offering mentorship or inviting women to contribute to specific activities. Another action that can be be adopted toward multiple strategies is to \textsc{create and enforce a code of conduct} by providing online training on enforcement and being transparent about the punishments for those who violate the code of conduct. There can be a training for contributors in general, another for mentors to (\textsc{prepare mentors to guide women)}, and a third for \review{allies} to advocate for women and act as ``collaborators, accomplices, and co-conspirators''~\cite{melaku2020better}. The content of the training can include practical examples of acceptable and non-acceptable behaviors. Communities can use mining tools to identifying gender pronouns in messages of mailing lists, pull-requests, and code reviews and help to \textsc{promote inclusive language}.% The authors of those messages can be individually contacted and even receive the non-compliance punishment.

\subsection{Research Opportunities} 

This subsection discusses the gaps in the literature that may be explored in future research.

\review{\noindent\textbf{Evaluation of recommended strategies to increase women's participation: }}
Although several strategies to increase women's participation have been proposed in the literature~\cite{PS3,PS6,PS8,PS9,PS12,PS16,PS23,PS24,PS25,PS26,PS27,PS28,PS29,PS30,PS31,PS32,PS33,PS34,PS35,PS36,PS37,PS38,PS39,PS40,PS41,PS42,PS47,PS48}, few works present scientific evidence about their effectiveness. For instance, Tourani et al.~\cite{PS33}, Imtiaz et al.~\cite{PS3}, and Vandana and Brandon~\cite{PS47} relegate the evaluation of the effectiveness of the ``code of conduct'' to future research, this is despite the fact that it is one of the most cited strategies to promote women participation. Izquierdo et al. \cite{PS12} discuss the difficulty of evaluating the effectiveness of strategies, as communities need to have consistent measurements before (baseline), during, and after their implementation. The authors reported that although OpenStack created the Women of OpenStack Working Group (which included educational sessions, professional networking, mentorship, social inclusion, and enhanced resource access), the OpenStack Foundation lacked baseline information about the involvement of women.

\review{\noindent\textbf{Theories to explain why women leave or avoid OSS projects:}}
The literature also reports a diverse set of challenges faced by women \cite{PS13,PS18,PS20,PS21,PS22,PS23,PS24,PS25,PS26,PS27,PS28}, but few make a theoretical connection to why women leave (or avoid) OSS projects.
%many of them without associated strategies or further understanding. 
Theoretical understandings can help create more effective, longer term solutions.
Some studies have analyzed motivation to participate in OSS projects~\cite{lakhani2003hackers,ghosh2002free,harsworking,hertel2003motivation}, but only a few report women's motivation, and none go deep in this analysis. Moreover, the literature lacks research exploring why women leave OSS, their motivation to avoid participating in OSS, and why a large portion of women who study STEM do not join OSS projects.
One of the few studies that have used theory to explain these phenomena is
Qui et al. \cite{PS46}; they found that social capital can support the long-term engagement for both men and women in OSS projects, and that when team members have more diverse programming language backgrounds, women are less likely to leave the project early. 
%Besides this study, there are no studies presenting theories to explain why women leave or do not join OSS projects. 

%For instance, non-inclusive communication issues can cause women to leave OSS \cite{PS13}, but there

\review{\noindent\textbf{Applying the method to study other minorities (in different domains):}} The lack of diversity affects different OSS and other domains in STEM. While in this study we focus on women in OSS, researchers can leverage this study's structure to investigate how the literature is positioned regarding the participation of other minority populations in OSS or even different domains. 
%The low frequency of women's involvement is explored in literature, but not the reasons for it. %Researchers can run empirical studies to implement and evaluate the proposed strategies and explore the motivations behind women's decisions to participate or not, and also their choices to leave an OSS project after they start contributing.

\subsection{Implications for Practice}
%In this section, we summarize the implications that this literature review brings to practice, as presented below.

\review{\noindent\textbf{Encouragement for women newcomers:}} Women newcomers can become aware of the different types of contributions made by other women through our study. Although not restricted to the activities presented here, women can be inspired by other women's success and motivations to participate, gain awareness of the challenges reported by other women so they can be prepared to face similar ones, and prioritize participation on projects that follow one or more reported strategies.

\review{\noindent\textbf{Opportunities for OSS communities:}} OSS communities that seek to foster gender diversity can combine and implement the strategies to increase women's participation, as we presented in the Discussion section. Moreover, being aware of the characteristics of women contributors, the communities can target their efforts on them by, for example, promoting events at universities (as most of the contributors are at least undergraduate students).

\review{\noindent\textbf{Opportunities for education:}} Being aware of the challenges that women face, educators can address the underlying issues causing these challenges in the classroom to improve students' (all gender) awareness of biases and discuss possible mitigation actions. This research can also inform educators who adopt contributions to OSS projects as a medium to teach software engineering~\cite{pinto2017training}.

%For example, Canedo et al. \cite{PS6} presented suggestions of actions that were provided by practitioners (women core developers), cited some cases where the actions are already happening, and suggested that combining strategies should contribute to women’s confidence to contribute to OSS communities, but did not provided empirical evidence about the effectiveness of them. Regarding the code of conduct, Robson~\cite{PS34} showed no significant difference in the proportion of women contributing to projects with or without a code of conduct, and concluded it would make litter difference to have a code of conduct unless it is enforced among the project members. Tourani et al.~\cite{PS33} presented cases where the code of conduct is enforced, but authors (and also Imtiaz et al.~\cite{PS3} and Vandana and Brandon~\cite{PS47}) called for further research to evaluate the effectiveness and best practices of adopting a code of conduct.

% Intel and Bitergia conducted research to produce the baseline and initial measurement of participation of different genders making different types of contributions, and then shared the results with the OpenStack community between 2016 and 2018, which were well-accepted by the community. 

%\section{Implications}

\section{Related Studies}

In this section, we present secondary studies about challenges and strategies to increase women's participation in STEM and in software engineering in particular.

\noindent\textbf{Women's participation in other cultures and domains.} Analogous to OSS projects, social and cultural context also influences women's barriers in the medical profession and their ability to rise to leadership positions~\cite{ramakrishnan2014women}. Women have played a significant and active role in many contemporary armed rebellions (where men are often presumed to be the default gender), and even are frequently involved in leadership roles~\cite{henshaw2016women}. By analyzing the career trajectories of women executives across a variety of sectors, Glass and Cook~\cite{glass2016leading} concluded that while attaining promotion to leadership is not easy, serving in a high position can be even more challenging. Although women can be more likely than men to be empowered to high-risk leadership positions, they often lack the support or authority to accomplish their strategic goals. As a result, women leaders often experience shorter tenures compared to men peers~\cite{glass2016leading}. Similar to software development teams, where women are instrumental to reducing community smells~\cite{catolino2019gender}, in international relations, the collaboration between women delegates and women civil society groups positively impacts and brings more durable peace when negotiating peace agreements~\cite{krause2018women}. The challenge of \textsc{Work-Life Balance} that we presented in Section~\ref{sec:rq6} is a general challenge faced by women who aim to work in Japan, where the low numbers of women in medicine reflect the prevalent societal belief that careers and motherhood do not mix~\cite{ramakrishnan2014women}. In contrast, Scandinavia has similar numbers of men and women physicians, which has coincided with the emergence of progressive work–life policies, the belief that women can combine motherhood and employment, and changing expectations of work-life balance. Historically, Sweden was the first country to establish paid parental leave for fathers in 1974, and its National Labor Market Board has developed statements since 1977 encouraging men to contribute to childcare responsibilities~\cite{haas2009fatherhood}.

\noindent\textbf{Challenges faced by women and strategies to increase women's participation in STEM.} The literature shows that the gender gap is largely present in the science, technology, engineering, and mathematics (STEM) fields at all education levels and in the labor market~\cite{fatourou2019women}. The barriers faced by women in STEM fields, presented by McCullough~\cite{mccullough2011women}, are similar to some of the challenges we presented in Section \ref{sec:rq6}, including discrimination and implicit bias (\textsc{Toxic Culture}), lifestyle choices, and family obligations (\textsc{Work-Life Balance Issues}), and lack of role models and mentors (\textsc{Community Reception Issues}).

%Ford et al.\cite{ford2017someone} identified that women who use Stack Overflow are more likely to contribute when they find other women contributing in the same community, which is similar to the kinship motivation and peer parity.

%\textbf{Strategies to increase women's participation in STEM.} 
Regarding strategies to increase women's participation, the W-STEM project~\cite{garcia2019engaging} seeks to create mechanisms to attract and guide women in Latin America in STEM higher education programs, including actions to monitor gender equality in enrollment and retention (as we presented in the strategy \textsc{promote awareness of the presence of peers}). Another strategy proposed by both Garcia-Holgado et al.~\cite{garcia2019engaging} and Moreno et al.'s~\cite{moreno2014women} study is to disseminate scientific and technological culture from an early age, promoting STEM studies vocation and choice for girls and young women in secondary schools (as we presented in the strategy \textsc{Promote women-specific groups and events}). Moreover, the women who participated on Moreno et al.'s~\cite{moreno2014women}'s study corroborated the strategy of \textsc{promote inclusive language} and the need to avoid chauvinistic attitudes, and to \textsc{de-stereotype} the CS student, suggesting that the general image of the student majoring in CS should change and not ascribe to the nerd stereotype, which is typically ascribed to men.

We also found secondary studies related to diversity in other domains. \citet{kazmi2014women,maheshwari2021review} show that women in academia and industry, despite their impressive performance, face work-life balance issues and the impostor syndrome \cite{kazmi2014women}, but can be supported by mentorship to overcome the challenges and advance in their career \cite{maheshwari2021review}. 

More specifically for Computer Science, \citet{pantic2019factors} found that individual (pre-arrival), institutional, and societal factors interplay to affect women's commitment and retention to the Computer Science program. \citet{felizardo2021global} found that research contributions from women in secondary studies have globally increased over the years, but are still concentrated in European countries. Although their findings are similar in terms of analyzing the factors that impact women's participation, none of them addressed specific motivations or challenges, or focused on strategies to increase women's participation in OSS.

\noindent\textbf{Other literature surveys about gender diversity in software engineering.} Gender (with a focus on women) is the most explored aspect of diversity in software engineering literature~\cite{menezes2018diversity,silveira2019systematic}. Spichkova et al.~\cite{spichkova2017role} analyzed the role of women within software architecture literature and found that it is understudied. In a more general literature review in Software Engineering, \citet{rodriguez2021perceived} highlighted that researchers have been exploring the gender bias problems in software engineering more than presenting the solutions to mitigate these problems. Our study complements these studies as it focuses on analyzing the literature to identify strategies proposed to mitigate the challenges faced by women in OSS.

The study closest to our work is a systematic literature review (SLR) conducted by Canedo et al.~\cite{dias2019barriers}. While their study investigated the problems causing women’s lack of engagement in software development in general, our analysis focused on the challenges faced by women in OSS projects (Section~\ref{sec:rq6}). Our study covered a larger spectrum of studies (51) as compared to Canedo et al. (24). Additionally, some of the studies used by Canedo et al.~\cite{dias2019barriers} did not qualify in our inclusion criteria, as these analyzed women participation in STEM in general. %We selected this paper as one of the primary studies due to another part of the paper that included an analysis of GSoC participants. The study indicates that women are underrepresented in the OSS community with less than 10\% of the total developers. 
 %We also consolidated and compared the motivations reported by literature as more relevant for men, for women, and for both genders.
When comparing the results, only two challenges were common between our work and Canedo et al.~\cite{dias2019barriers}---Toxic Culture and Impostor Syndrome. 
%We complemented the challenges reported by primary studies including Lack of Peer Parity, Non-Inclusive Communication, Stereotyping, Community Reception Issues, and Work-Life Balance Issues. 
Canedo et al.~\cite{dias2019barriers} also presented possible solutions to increase the engagement of women in OSS. Our work expands this list (Promote inclusive language, Encourage and be welcoming to women, \review{de-bias} tools, Recognize women’s achievement (visibility) and Prepare mentors to guide women). Finally, in our work we analyzed the motivations that drive women to contribute (Section~\ref{sec:rq5}) and further analyzed and associated the challenges faced by women in OSS (Section~\ref{sec:rq6}) with the strategies to mitigate those challenges (Section~\ref{sec:rq7}) as proposed in the primary studies (Fig.~\ref{strategies_challenges}).

%Furthermore, we investigated why Canedo et al.~\cite{PS43} SLR identified 24 qualified papers whereas only 15 of those papers were part of our 51 qualified papers. 

%\input{Tables/comparison_canedo}

%To the best of our knowledge, there are no other studies aggregating evidence about women's participation in OSS projects.
Considering that women are still underrepresented in OSS and projects seek more diversity to increase productivity and enhance the quality of delivered software, it is crucial to understand the profile of women who contribute to OSS, including their motivations and challenges, as well as strategies to attract and retain them. Therefore, we believe that the consolidated relationships presented in this literature review are necessary and could inform both practitioners and researchers.

\section{Threats to Validity}
We used the checklist provided by Ampatzogloua et al.~\cite{ampatzoglou2019identifying} to identify and categorize threats to validity and corresponding mitigation actions.

\textbf{Study inclusion/exclusion bias.}
%This study has two fundamental threats to sampling adequacy. 
There may be relevant papers in the population that are not in our sampling frame, e.g., studies that covered women's participation in OSS published in outlets that are not indexed by the target search engines or excluded by the filters we used (see Section~\ref{selection_criteria}). Also, there may be relevant articles in the sampling frame that are not in the sample; for example, there may be papers covering women's participation in OSS never use the terms we included in the search string. We took several steps to mitigate these threats, including using additional terms besides ``women'' in the search string (e.g., ``diversity'' and ``heterogeneous team''), backward and forward snowballing, and having the list of primary studies reviewed by the most prolific authors. Although in this paper we use the term ``women'' and ``men'' as a shorthand for people who self-identify as such, including all dimensions of gender, such as transgender and non-binary identifications, there can be dissimilar participation of those from the different dimensions of women that were not reported by the primary studies.

\textbf{Researcher bias and Repeatability.}
We used an open research question (and search string) to find all forms of participation by women in OSS. This openness can bring subjectivity to the data extraction and repeatability. While the first author of this study led the execution of steps to filter and select the primary studies (see Fig~\ref{fig_steps_select_primary_studies}, the other two researchers with extensive experience in conducting qualitative analysis and literature studies participated in defining the protocol with phases and steps (see Fig~\ref{fig_method}), inclusion and exclusion criteria, and discussed the data in weekly meetings during the recording process. Additionally, one senior researcher cross-checked a portion of the extracted data. During the weekly meetings, we discussed and adjusted codes and categories until we reached agreement. In the meetings, we also checked the consistency of our interpretations. %All analysis was thoroughly grounded in the data collected from the primary papers, and exhaustively discussed amongst the whole team to reach an agreement.

\textbf{Robustness of classification.}
We used existing classification schema from the literature to explain two categories of participation: the types of contributions that women make in OSS projects (see Section~\ref{sec:rq3}) and the motivations that drive women to participate in OSS projects (see Section~\ref{sec:rq5}). The selection of those two initial
classification schema poses a threat, since they might not be enough to explain the phenomena. At the same time that we had these two schemas to guide our analysis, we constantly assessed the possibility and need to extend them. Therefore, although we did not find any new category of motivation to extend Von Krogh et al.'s \cite{von2012carrots} framework of OSS motivations, we extended the classification used for types of contributions with new roles that arose from the primary studies. 

%To increase consistency, we performed data analysis in a group of three researchers. We had weekly meetings to discuss and adjust codes and categories until we reached agreement. In the meetings, we also checked the consistency of our interpretations. All analysis was thoroughly grounded in the data collected from the primary papers, and exhaustively discussed amongst the whole team to reach an agreement. Our research team also contain senior researchers with extensive experience in conducting qualitative analysis and systematic literature studies, who also examined the papers' selection and interpreted the extracted results to mitigate inclusion, exclusion, and researcher bias.

%Different studies in different time frames may pose different results, as women's participation can change over the years. To avoid misinterpretation, we reported the results bringing the context of each primary study.

%We included terms in the search string (e.g., diversity and heterogeneous team) that returned a large number of primary studies (including many irrelevant ones). From 1914 unique papers, we considered only 155 (8\%) in the first filter and 35 (2\%) after full-reading.

\section{Conclusion}

In this paper, we investigated women's participation in OSS projects in terms of frequency and demographics, their motivations to participate, the types of contributions women make, the challenges they face, and the proposed strategies to mitigate those challenges.

Our literature mapping has shown severe gender disparity in OSS, with women's representation being at about 10\%. Further, the women who participate are generally volunteers who can devote less than one workday a week to OSS.
%Our findings show that, according to the literature, women represent about 10\% of Open Source Software contributors, considering different types of participation. The majority of women contribute to OSS projects for less than five years, devote less than one work-day to OSS on few repositories, and are at least undergraduate students. 
Although present in community-centric roles, women are less likely to be authors of pull requests and core developers, with many making non-code contributions.
Gender biases exist in OSS in several places. For example, when submitting a pull request, women have high rates of merge acceptance, but lower when they explicitly identify themselves as women. 
Women also face social challenges such as a lack of peer parity, non-inclusive communication, a toxic culture, impostor syndrome, community reception issues, stereotyping, work-life balance issues, and bias against gender-biased peer review.
%Reciprocity, kinship and getting paid to contribute motivates more women than men. Altruism, learning, and own-use motivate both men and women.
%, career and reputation are low-ranked for both genders. 
OSS communities that seek to increase women's participation can mitigate these challenges by providing awareness about the presence of other women, promoting inclusive language, organizing women-specific groups and events, de-stereotyping the OSS contributor, encouraging and welcoming women, placing women in leadership, adopting de-biasing tools, recognizing women's achievement, preparing mentors to guide women, and creating and enforcing a code of conduct.

As future steps, we plan an in-depth analysis of women's motivations, challenges, and reasons that would make them leave. We will propose strategies to attract and retain more women to OSS communities, as a long-term plan.

%Our goal with this research is to increase diversity in OSS by helping minorities to join, be recognized, and part of the core of OSS projects.

\section{Acknowledgements}
This work is partially supported by the National Science Foundation under Grant Numbers 1815486, 1815503, 1900903, and 1901031.

\bibliographystyle{ACM-Reference-Format}
\bibliography{references}

%\ifCLASSOPTIONcaptionsoff
%\newpage
%\fi

\clearpage
\newpage

%\noindent{\large\bfseries Appendix A.\par}

\nocitesec{*}
\bibliographystylesec{appendixStyle}
\bibliographysec{primaryStudies}

%\printbibliography[title={Appendix references}]

\end{document}